\documentclass[a4paper,twocolumn,11pt,accepted=2024-01-03]{quantumarticle}
\pdfoutput=1
\usepackage[utf8]{inputenc}
\usepackage[english]{babel}
\usepackage[T1]{fontenc}
\usepackage{amsmath}
\usepackage{hyperref}
\usepackage{tikz}
\usepackage{lipsum}

\usepackage{tikz}
\usetikzlibrary{calc}
\usepackage{amsmath}
\usepackage{graphicx}
\usepackage{amsfonts}

\newcommand{\tr}{\mbox{tr}}
\usepackage{amssymb}
\usepackage{amstext}
\newcommand{\ignore}[1]{}

\newcommand{\ket}[1]{| #1 \rangle}
\newcommand{\bra}[1]{\langle #1 |}
\newcommand{\cyl}[1]{{\mathcal{M}}^* (#1)}
\newcommand{\dua}{{\mathcal{M}}^*}

\usepackage{soul}%allows highlighting

\usepackage{ltxgrid}  %  widetext 

\begin{document}

%%%%%%%%%%%%%%%%%%%%%%%%%%%%%%%%%%%%%%%%%%%%%%%%%%%%%%%%%%%%%%%%%%%%%%%%%%%%%%%
\title{Efficient classical simulation of cluster state quantum circuits with alternative inputs}
% AUTHOR
\author{Sahar Atallah}
\affiliation{Department of Mathematics, Brunel University London, Kingston Ln, Uxbridge, UB8 3PH, United Kingdom}
% AUTHOR
\author{Michael Garn}
\email{Michael.Garn@brunel.ac.uk}
%\orcid{0000-0003-0290-4698}
% AUTHOR
\affiliation{Department of Mathematics, Brunel University London, Kingston Ln, Uxbridge, UB8 3PH, United Kingdom}
\author{Sania Jevtic}
\affiliation{Phytoform Labs Ltd., Lawes Open Innovation Hub, West Common, Harpenden, Hertfordshire, England, AL5 2JQ, United Kingdom}
%\orcid{0000-0003-1985-4623}
% AUTHOR
\author{Yukuan Tao}
\email{Yukuan.tao.gr@dartmouth.edu}
\affiliation{Department of Physics and Astronomy, Dartmouth College, Hanover, New Hampshire, 03755, USA}
% AUTHOR
\author{Shashank Virmani}
\email{Shash.Virmani@brunel.ac.uk}
\affiliation{Department of Mathematics, Brunel University London, Kingston Ln, Uxbridge, UB8 3PH, United Kingdom}
%\orcid{0000-0003-1533-8015}

%%%%%%%%%%%%%%%%%%%%%%%%%%%%%%%%%%%%%%%%%%%%%%%%%%%%%%%%%%%%%%%%%%%%%%%%%%%%%%%
\maketitle
%%%%%%%%%%%%%%%%%%%%%%%%%%%%%%%%%%%%%%%%%%%%%%%%%%%%%%%%%%%%%%%%%%%%%%%%%%%%%%%
\begin{abstract}
 We provide new examples of pure entangled systems related to cluster state quantum computation that can be efficiently simulated classically. In cluster state quantum computation input qubits are initialised in the `equator' of the Bloch sphere, $CZ$ gates are applied, and finally the qubits are measured adaptively using $Z$ measurements or
measurements of $\cos(\theta)X + \sin(\theta)Y$ operators. We consider what happens when the initialisation step is modified, and show that for lattices of finite degree $D$, there is a constant $\lambda \approx 2.06$ such that if each individual qubit is prepared in a state that is within $\lambda^{-D}$ in trace distance of a state that is diagonal in the computational basis, then the system can be efficiently simulated classically in the sense of sampling from the output distribution within a desired total variation distance. In the square lattice with $D=4$ for instance, $\lambda^{-D} \approx 0.056$. We develop a coarse grained version of the argument which increases the size of the classically efficient region. In the case of the square lattice of qubits, the size of the classically simulatable region increases in size to at least around $\approx 0.070$, and in fact probably increases to around $\approx 0.1$. The results generalise to a broader family of systems, including qudit systems where the interaction is diagonal in the computational basis and the measurements are either in the computational basis or unbiased to it. Potential readers who only want the short version can get much of the intuition from figures 1 to 3.
\end{abstract}
%%%%%%%%%%%%%%%%%%%%%%%%%%%%%%%%%%%%%%%%%%%%%%%%%%%%%%%%%%%%%%%%%%%%%%%%%%%%%%%
\section{Introduction and summary of main results}

An important open problem in quantum computing is to understand when quantum systems  can or cannot be efficiently simulated classically. The observation that classical simulation methods fail to be efficient for generic quantum systems was the original motivation for quantum computation \cite{Preskill}. While there are rigorous proofs of quantum computational advantage in certain settings such as communication complexity or with depth restrictions \cite{Raz,tomamichel2019quantum}, it is still in principle possible (however unlikely) that without such restrictions quantum computers can be efficiently simulated classically. The question has been given added impetus recently, as it is central to the discussion surrounding recent quantum supremacy experiments \cite{Google,Pan,napp2019efficient,noh2020efficient,AharonovGao}. 

Classical algorithms for simulating quantum systems often have a wide variety of aims, such as computing probabilities, estimating physical quantities in many-body systems, sampling observed distributions, or theoretical investigation of the sources of quantum advantage. Furthermore, different definitions of the phrase ``efficient classical simulation" are often used (see \cite{Hakop} for a recent discussion of various possibilities). However, in spite of this diversity of aims and motivations, certain themes repeatedly arise. This is because the classical algorithms that have been proposed often work by (a) singling out a special feature of quantum theory as one of the possible sources of non-classicality, and then (b) simulating systems for which that feature is limited. 

Some of the earliest examples of this approach highlighted quantum entanglement as the `special feature'. It was shown that in some settings a lack of quantum entanglement (perhaps due to noise) can lead to efficient classical simulation algorithms \cite{Jozsa,ABnoisy,HN,Yogancade}. Even if entanglement is present, but its structure is limited to being of low `width' or `tree-width', then methods with a tensor-network flavour can often be exploited to provide efficient classical simulations \cite{Nielsen1D, Yoran, markov, jozsa2006simulation,van2007classical,Hastings, Ghosh}. In the context of many-body physics, such limited entanglement may be exploited to compute physically important quantities \cite{Vidal,peps}. 

However, it is by now well known that entanglement is far from the full story. Seemingly weak amounts of entanglement can in fact be strong enough for quantum computation \cite{van2013universal,gross2007measurement}, and seemingly large amounts of quantum entanglement can be efficiently simulated classically. The well known Gottesman-Knill theorem \cite{GK}, for example, shows that stabilizer computations - which can demonstrate highly non-classical features such as non-locality - can be efficiently simulated classically without some form of additional `magic' \cite{KnillMagic,BK}. This idea has been expanded upon in several works (e.g. \cite{nest2008classical,jozsa2013classical}), and has applications such as providing classical simulations of systems with a small amount of `magic' \cite{BravyiRank1,BravyiRank2,pashayan2021fast,seddon2021quantifying,qassim2021improved}, as well as providing upper bounds to fault tolerance thresholds \cite{VHP05,Buhrman}.
Through the formalism of discrete Wigner functions there are also connections between the stabilizer formalism and quasi-probability distributions that arise in attempts to give quantum theory an alternative `realistic' description \cite{Galvao,GrossPhase1,GrossPhase2}. A number of other (sometimes efficient) classical descriptions of quantum computation based on quasi-probability distributions or non-quantum operators have also been developed, e.g. \cite{Galvao,Pashayan2015WB,raussendorf2020phase,okay2021extremal,zurel2020hidden,RV,RVprep,AJRV1}. Our categorisation of the above classical algorithms into various themes is not objective, there are often connections between them, and moreover there are improvements or further insights to be gained by combining different approaches (e.g. \cite{gosset2020fast} combines Gottesman-Knill with tensor network methods, and \cite{schwarz2013simulating} building upon \cite{nest2011classical} shows that quantum advantage requires output distributions to be not too sparse). 

Another important class of classically efficiently simulatable quantum systems arises from Valiant's matchgate algorithm and extensions \cite{Valiant,Terhalferm,jozsa2008matchgates,Brod}. These algorithms provide efficient classical simulations of entangled systems that are strongly related to fermionic physics \cite{Terhalferm}. The work of \cite{Somma} generalised some of these ideas to what the authors termed {\it Lie Algebraic} models of computing (`LQC'). One of the themes of that work is an idea that the notion of entanglement can be generalised \cite{Barnum1,Barnum2,Barnum3} to a new notion which is defined relative to a privileged set of observables, and in some situations this can be exploited to give classical simulation methods. A different type of generalised entanglement (although it is still an instance of the broad framework put forward in \cite{Barnum1,Barnum2,Barnum3}) was utilised by some of us in \cite{RV,RVprep,AJRV1} to develop local hidden variable models and classical simulation methods in other situations. Our present work will be in a similar vein to these works.

In this work we will exploit a version of generalised entanglement (or more precisely generalised separability) to provide new examples of (pure) multiparty entangled systems that can be efficiently simulated classically. All our results concern variations on cluster state quantum computing, and are summarised as follows:

\begin{enumerate}
    \item We will begin by considering what happens when we vary the inputs of qubit cluster state quantum computing \cite{Raussendorf} architectures, keeping the permitted measurements the same. We will find that when each input qubit is initialised not in the standard $\ket{+}$ states, but in single qubit states that are sufficiently close to a state that is diagonal in the computational basis, then the systems can be efficiently simulated classically in the sense of sampling the outcomes to arbitrary additive error in polynomial time (see footnote {\footnote{This means sampling in polynomial time from a probability distribution $p$ that satisfies $\|p-q\| \leq \epsilon$ where $q$ is the probability distribution of measurement outcomes on the system, $\epsilon > 0$ is an arbitrary fixed constant, and the norm is the total variation distance}} for a precise definition of this.). This is in spite of the fact that the resulting systems (after the $CZ$ interactions have been applied) include pure multiparty entangled states that contain a large amount of `magic' \cite{KnillMagic,BK}, and retain several of the important features present in cluster state quantum computing (except for a particular form of non-locality - see later discussion). The key technical idea is a lemma showing that the control-$Z$ ($CZ$) interactions do not lead to a particular generalised form of entanglement. For our purposes this means the following: in the usual expression $\sum_i p_i \rho^A_i \otimes \rho^B_i$ for quantum separable states the local operators $\rho^A_i,\rho^B_i$ must be quantum states, but we will relax this, allowing the local operators $\rho^A_i, \rho^B_i$ to come from more general sets than the local quantum states, e.g. allowing some operators with negative eigenvalues. We will later see that if are careful to control this `negativity', we may exploit generalised separable decompositions, together with an existing method \cite{HN} for non-entangled quantum systems, to efficiently simulate certain input states. In particular, defining the radius $r$ of a unit trace $2 \times 2$ Hermitian matrix $\rho$ as $r := \|\rho-\rho_{diag}\|$, where the norm is the trace norm and $\rho_{diag}$ is obtained from $\rho$ by setting off-diagonal elements to zero, our algorithm classically simulates efficiently when the each input qubit has $r  \leq \lambda^{-D}$, where $\lambda \approx 2.06$ and $D$ is the maximum degree of the underlying graph. Note that the inputs can include non-quantum operators with negative eigenvalues, however it turns out that they do not lead to negative probabilities when used as inputs for the cluster state circuits that we consider.
\item We then generalise the approach to measurement based quantum computation in what we term {\it privileged basis system} measurement based quantum computation (``PBS" for short). While we define these systems precisely later, they are measurement based quantum computations in which the computational resource state (i.e. the analogue of the cluster state) is created by diagonal (in the computational basis) unitaries acting upon input particles placed on a lattice, and the allowed measurements are restricted in a particular way. This class of systems include the original cluster state scheme, as well as a number of other MBQC proposals (e.g. \cite{gross2007measurement,KissingerW,Tomo,MillerM1,MillerM2, Hall}). For any such systems our results imply that there is an analogue of $\lambda^{-1}$, i.e. there is a constant $c$ such that if each individual input particle is initialised from within a particular set of `size' (a term whose meaning will become clear later) $c^D$ around the diagonal single particle states, then the systems can be efficiently simulated classically.

We remark that the results demonstrate any PBS that is capable of non-classical computation for at least some input states (examples include \cite{gross2007measurement,KissingerW,Tomo,MillerM1,MillerM2,Hall}) has at least one non-trivial ``computational transition": for all such systems thin enough starting `cylinders' provide a finite size convex region of the single particle input states, including pure inputs, that can be efficiently simulated classically, even though for other inputs non-classical computation is possible.

\item Borrowing a common paradigm from many-body physics, we then explore a `coarse grained' version of the simulation method for sufficiently regular lattices. By this we mean the following: we cut the system into blocks of qubits, treat each block as a single particle on a new lattice, and then construct a decomposition over these blocks that does not exhibit a particular generalised form of entanglement. It turns out that this process leads to classical simulation algorithms for an increased set of inputs, as well as some interesting mathematical structure. In particular, given a suitable lattice we parameterise the size of each block by a positive integer $n$ (the details of which we describe in section \ref{SectionCoarse}), and we find that this leads to two convergent sequences $l_n$ and $u_n$ with the following properties:
\begin{enumerate}
\item $u_n$ and $l_n$ are the solutions to two families of optimisation problems, which are related to each other by a change of parameters,
    \item $u_n$ is non-increasing, $l_n$ is non-decreasing, and $u_n \geq l_n$,
    \item Input single particle states (of the physical particles, not the blocks) with radius $r < l := \lim_{n \rightarrow \infty} l_n$ can be efficiently simulated classically,
    \item For inputs with radius $r > u := \lim_{n \rightarrow \infty} u_n$, the particular notion of generalised separability that we use for our coarse graining approach breaks down, leading to an ill-defined sampling problem. The details of this discussion are left until later.
\end{enumerate}
In the case of the square 2D lattice and $CZ$ interactions we have numerically computed upper bounds to $u$ and lower bounds to $l$ and are quite certain that $0.0698 \leq l \leq u \leq 0.139$, but based on a conjecture and small scale numerical experiments we expect that in fact $0.0913 \leq l \leq u \leq 0.128$. These values should be compared to $\lambda^{-4} \approx 0.056$, which would be the size of classically simulatable region without coarse graining.
\end{enumerate}

%%%%%%%%%%%%%%%%%%%%%%%%%%%%%%%%%%%%%%%%%%%%%%%%%%%%%%%%%%%%%%%%%%%%%%%%%%%%%%%

\section{Prior Work and Context}

A cluster state computation \cite{Raussendorf} in its original form proceeds by placing $\ket{\psi} = \ket{+} = (\ket{0}+\ket{1})/\sqrt{2}$ states on the vertices of a graph, interacting neighbouring qubits with control-$Z$ (henceforth denoted as $CZ$) gates, and then destructively measuring (i.e. measured qubits are not reused) in the $Z$ basis or the $XY$ plane (i.e. measurements of operators of the form $\cos(\theta) X + \sin(\theta) Y$). This has the power of BQP. 

 What happens if $\ket{\psi}$ is replaced by another pure or mixed state state $\rho$? Two facts are immediate from the original scheme \cite{Raussendorf}: if $\rho$ corresponds to an equal weight superposition of $\ket{0}$ and $\ket{1}$, then the power remains that of BQP (this is because the computational power is trivially invariant under rotations about the $Z$ axis), and if $\rho=\ket{0}\bra{0}$ or $\rho=\ket{1}\bra{1}$, then the system can be efficiently simulated classically as the $CZ$ gates act trivially and so the final state is a product state. 
 
 Previous works have looked at other input states, although sometimes in slightly different settings to the one considered in this work. In \cite{Terry,mora_universal_2010} for example, all local measurements are permitted, and moreover the measurements are permitted to be nondestructive (in this work we will only consider destructive measurements). Nevertheless, in that setting it was shown that for some graphs when $\rho$ is a pure or mixed state close enough to $\ket{+}$, quantum computation is still possible by performing filtering measurements that probabilistically distill out a perfect cluster state (for some graphs it is easy to adapt the approach to destructive measurements of the original cluster state form - we discuss this briefly in section \ref{SectionObstacles}). It was also shown that when there is sufficient noise - enough to prevent large scale quantum entanglement in a given graph - the systems can be efficiently simulated classically. Such considerations are also present in \cite{Dan}, albeit for a quite different noise model. The core idea that noise can destroy computationally useful entanglement goes back to the early years of quantum computing, see e.g. \cite{ABnoisy}.

In the case of sufficiently noisy input states, other methods can also be used to provide classically efficiently simulatable regimes. For instance, enough dephasing will turn each $CZ$ gate in the cluster state circuit into one that does not generate quantum entanglement, thereby allowing the classical algorithm of \cite{HN} to be used. Alternatively, dephasing an ideal input $\ket{+}$ will effectively (by shifting the noise through to the measurements) turn the measurements into Clifford ones so that the Gottesman-Knill theorem \cite{GK} may be invoked. For any underlying graph if $\rho$ is a dephased $\ket{+}$ state, then the Gottesman-Knill theorem classically simulates once $\| \rho - \rho_{diag}\| \lessapprox 0.7$ (see e.g. \cite{VHP05}).

However, all of these previous approaches need the inputs to be mixed or noisy in order to enter a classically efficient regime. This is where our work is differs most from previous literature: apart from trivial $\ket{0},\ket{1}$ inputs, or for systems with suitably restricted connectivity (such 1D, low width, or low tree-width systems
\cite{Nielsen1D, Yoran, markov, jozsa2006simulation}, or when qubit loss significantly limits the size of clusters \cite{Dan}), previous works have required non-zero quantum entropy to bring on a classically efficient regime in the kinds of systems that we consider. In contrast, in this work we develop classically efficient simulation algorithms in which the single qubit initialisations $\ket{\psi}$ can be taken to be both {\it pure} or mixed states as long they are close enough to diagonal in the computational basis.

In order to achieve this we only allow the original cluster state measurements (i.e. {\it destructive} measurements $Z$ basis and $XY$ plane measurements). However, these measurements are still non-trivial because when the inputs are the ideal $\ket{+}$ states they are sufficient for quantum computation. To our knowledge no previous classical algorithm efficiently simulates the pure systems that our method can efficiently simulate. Moreover, the method we develop has a natural generalisation to a wide variety of other entangled states, and compared to most previous classical algorithms for these types of systems, our  approach (at least the non-coarse-grained version) is less reliant on specific features of the underlying graph (such as percolation thresholds or tree-width) as it only cares about the degree of the graph (the coarse grained version of our argument does rely more on the graph structure).

We note that while our methods are the only known efficient classical methods for the low entropy instances that we consider, for sufficiently noisy inputs with $CZ$ interactions an approach based on the Gottesman-Knill theorem is more powerful than the techniques presented in this work. The Gottesman-Knill theorem does not apply to our low entropy systems (because by magic state distillation \cite{BK}) they contain enough `magic' to enable quantum computation given access to arbitrary stabilizer computation), and furthermore the Gottesman-Knill theorem might not be effective when the $CZ$ gate is replaced by other non-Clifford diagonal gates - as we consider when generalising Lemma 1 in section \ref{section_generalisation}.

It is tempting to argue that results of the form that we develop in this article should be either expected or evident: one might argue that if input qubit state $\rho$ is close to diagonal, e.g. if $\rho$ is a pure state of the form $\ket{0}+\epsilon\ket{1}$, with $\epsilon$ small, then even after the $CZ$ gates there will be little entanglement, and so the system is likely to be classically efficiently simulatable \footnote{We note that simply approximating $(\ket{0}+\epsilon\ket{1})^{\otimes n}$ by $\ket{0}^{\otimes n}$ will not give a good simulation, as the overlap scales as $\epsilon^n \rightarrow 0$. Even though our measurements are restricted, the trivial case of a 2-colourable graph (such as the 2D square lattice) that is broken into disconnected single qubits by measuring one colour of qubits in the $Z$ basis, already shows that $\ket{0}^{\otimes n}$ gives a poor approximation even under the allowed measurements: if we measure the other colour in the $X$ basis, the approximation that they were initialised in $\ket{0}$ will give uniformly random outcomes, the exact initialisation $\ket{0}+\epsilon\ket{1}$ will rapidly show bias as $n$ increases.}. However, one has to be careful with this kind of reasoning for a few reasons. Firstly, states that are locally close to product states can still support measurement based quantum computation - see \cite{gross2007measurement} (similar statements are true  for the gate model \cite{van2013universal} too). Secondly, it is conceivable, although we do not yet have a proof of this, that if all destructive local measurements are allowed (as opposed to restricting the measurements to the standard cluster state measurements) then some of the pure systems that we demonstrate to be classically efficiently simulatable might gain the ability to perform universal quantum computation or some form of non-classical computation (in the sense that an efficient classical sampling to additive error is not possible). If this turns out to be the case, then it would rule out any classical simulation method that does not exploit the restriction on measurements, and the intuitive idea that the systems are weakly quantum entangled would be false (in this context, in section \ref{SectionObstacles} we point out that allowing all local destructive measurements indeed does bring an additional power on some lattices - that of being able to create ideal cluster states under postselection, even when the inputs are very close to $\ket{0}$ or $\ket{1}$). Finally, even if the intuition is true that the kind of quantum entanglement we have in our systems is too limited to allow non-classical computation, then one still has the challenge of working out how far from diagonal the inputs can be while remaining classically efficiently simulatable. In this respect our method is technically appealing in that it gives rigorous quantitative bounds applicable to any lattice.

One might speculate that some variant of a tensor network based method (e.g. \cite{van2007classical,Yoran,markov,jozsa2006simulation}) may be used to supply an efficient classical simulation of the systems that we consider. However, the entangled state resulting from inputs $\rho$ of the form $\ket{0}+\epsilon \ket{1}$ can be transformed to the ideal cluster state (as arises when $\rho$ is given by $\ket{+}$) by applying local linear transformations $A_{\epsilon} = \ket{0}\bra{0} + (1/\epsilon)\ket{1}\bra{1}$. As ideal cluster states are not likely to be classically efficiently simulatable, this suggests that any method working using some form of efficient tensor manipulation would need to exploit not just the tensor network structure, but some property resulting from the transformation $A_{\epsilon}$ - perhaps the fact that acting with $A^{-1}_{\epsilon} =  \ket{0}\bra{0} + \epsilon\ket{1}\bra{1}$ locally on each qubit of an ideal cluster state reduces correlations between different parts of the state. However, even if such an approach is possible (and it would certainly not be if allowing all local destructive measurements brought non-classical computational power), then we anticipate that it would probably need to exploit further structure in the underlying graph than just the degree, and would also likely be more technically challenging. This is because any classical simulation method is anticipated to fail for $\epsilon \approx 1$, and so any classical algorithm has to have a `phase transition' at which it fails. For our method we are able to rigorously identify classical regions without needing to invoke the typical kinds of technical machinery (e.g. percolation thresholds) that might be needed to identify a phase-transition. However, if a tensor network approach does turn out to work, it would have an advantage over our approach in that it would apply to all destructive (and possibly even non-destructive) local measurements, not just the cluster state measurements we consider here.

Our work is part of a sequence of papers in which some of us have investigated a specific notion of generalised entanglement (a particular instance of the more general notions considered in \cite{Barnum1,Barnum2,Barnum3}) to construct classical simulation algorithms and local hidden variable models \cite{RV,RVprep,AJRV1,AJRV2_1,AJRV2_2}. In \cite{AJRV1} a reasonably general construction is given that, given almost any set of restricted local measurements on local particles of high enough dimension ($\geq 16$), allows one to write down a $pure$ multiparticle entangled state that has a local hidden variable model for those measurements - something that would be impossible if all local measurements are allowed \cite{LoPop}. Moreover, some of those examples have the following property: they can be efficiently sampled classically by exploiting a type of generalised separability, but if all measurements are permitted they enable universal quantum computation, and hence no classical efficient simulation is likely unless it exploits the restricted measurements. While the examples of \cite{AJRV1} might be considered somewhat contrived as compared to the situations considered in this paper, together they demonstrate that there can be large scale, computationally significant, differences between generalised entanglement and the regular quantum version.

The version of generalised entanglement that we use should be contrasted to more general versions developed in \cite{Barnum1,Barnum2,Barnum3}, which may be briefly summarised as follows. In the conventional study of quantum entanglement one considers the comparison between a global system and subsystems, and global states are said to entangled if they cannot be described as convex mixtures of pure products of the individual subsystem states. In \cite{Barnum1,Barnum2,Barnum3} this perspective is modified to consider comparisons between one algebra of observables and a subalgebra, or one convex set and a subset, accompanied by an appropriate definition for states of the system. By considering subalgebras or subsets rather than subsystems, their notion of generalised entanglement does not necessarily need a partition of the system into subsystems, in contrast to both the usual notion of quantum entanglement and the particular version of generalised entanglement that we will consider in this work. In the context of Lie algebras the viewpoint of \cite{Barnum1,Barnum2,Barnum3} was adopted in \cite{Somma} to develop efficient classical algorithms for some quantum-entangled situations that generalise previously known fermionic \cite{TerhalD} systems. However, in spite of the fact that it also uses a notion of generalised entanglement to motivate a classical simulation algorithm, the Lie algebraic framework of \cite{Somma} does not apply to the situations we consider in this paper.

The results we obtain have relationships to other foundational questions. Our classical simulation algorithms provide two types of hidden variable model for entangled pure states - the first is a local hidden variable model in the conventional sense, and the second (resulting from the `coarse grained' version) is a local hidden variable model where particles can communicate within certain blocks. Our approach also can be considered as an instance of a type of non-quantum theory which has certain non-classical features (such as an uncertainty principle for some measurements) but no entanglement. It is not quite a generalised probabilistic theory in the sense considered in \cite{Boxworld1,Boxworld2,Boxworld3,Boxworld4}, because in some situations it can lead to negative probabilities. However, it fits into the broad theme of computation in beyond-quantum theories that has been explored in recent works \cite{Lee}.

%%%%%%%%%%%%%%%%%%%%%%%%%%%%%%%%%%%%%%%%%%%%%%%%%%%%%%%%%%%%%%%%%%%%%%%%%%%%%%%

\section{Cylindrical state spaces and preview of main techniques}

Let us first explain what we mean by generalised entanglement. As the version of generalised entanglement that we need here is a special case of the broad framework developed in \cite{Barnum1,Barnum2,Barnum3}, we will be more narrow and concrete than \cite{Barnum1,Barnum2,Barnum3} in our description.

One can consider modifying local state spaces so that they are not sets of quantum operators, but more general sets of operators that may only return valid probabilities under a restricted set of measurements of interest. Such new sets of operators can change our notion of entanglement, and this can lead to classical descriptions where they might otherwise be unexpected. Indeed, the well known fact that a Bell state such as $(\ket{00}+\ket{11})/\sqrt{2}$ has a local hidden variable model with respect to Pauli measurements can be reinterpreted as a statement that the state is separable w.r.t. cubes of operators \cite{RV} that arise in the study of discrete phase spaces \cite{GrossPhase1,GrossPhase2}.

\begin{figure}[ht!]
\centering
\begin{tikzpicture}
%\hspace{-1cm}
%\includegraphics[width=80mm]{cylinderpic.eps}

\draw[help lines, color=gray!30, dashed] (-3,-3) grid (3,3);
\draw[->,gray] (0,0)--(2.8,0) node[below]{$y$};
\draw[->,gray] (0,0)--(0,2.8) node[right]{$z$};
\draw[->,gray] (0,0)--(-2.8,-1) node[below]{$x$};

%CYLINDER
%vertical lines down
\draw (-2,2) -- (-2,-2);
\draw (2,2) -- (2,-2);
%top cylinder: ellipse
\draw (0,2) ellipse (2 and 0.5);
%arc at bottom & dashed arc
\draw (-2,-2) arc (180:360:2 and 0.5);
\draw [dashed] (-2,-2) arc (180:360:2 and -0.5);

%SPHERE
%ball colour, opacity = 'darkness': centered at 0,0 with size 2cm
\shade[ball color = gray!40, opacity = 0.5] (0,0) circle (2cm);
%draw outline/perimeter of circle 
\draw (0,0) circle (2cm);
% XY plane, starting at -2,0  
\draw (-2,0) arc (180:360:2 and 0.6);
% dashed line
\draw[dashed] (2,0) arc (0:180:2 and 0.6);
%node in middle
\fill[fill=black] (0,0) circle (2pt);

%dashed line of radius
\draw[->, dashed,thick] (0,2) -- node[above]{$r$} (2,2);
%\draw[->, dashed,thick] (0,2) -- node[above]{$r$} (-2,2);
%\draw[->, dashed,thick] (0,1) -- node[above]{$r$} (2,1);

%SOME DOTS
\fill[fill=black] (0,2) circle (2pt);
\fill[fill=black] (0,-2) circle (2pt);
%\draw[->,dashed, black,ultra thick] (0,2)--(2,2) node[above]{$r$} (1,2);
\end{tikzpicture}

\caption{This diagram illustrates a cylinder with $r=1$ in Bloch space. The sphere
represents the Bloch Sphere. Our cylinders always extend the full height from $z=-1$ to $z=+1$, irrespective of radius. The unit cylinder with $r=1$ is the normalised
dual of the permitted measurements (i.e. the set of normalised operators that yield positive probabilities for the allowed measurements).}
\label{cylinderpic}
\end{figure}
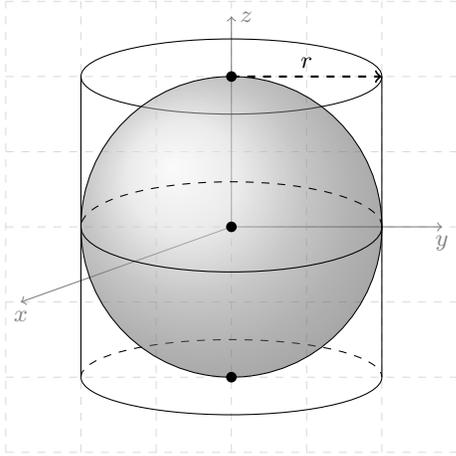

We will consider the action of $CZ$ (control-$Z$) gates. When a $CZ$ gate acts on input pure product states that are not computational basis states, the output will be quantum entangled, in that it cannot be given a quantum separable decomposition of the form:
\begin{equation} \label{sepfirst}
\sum_{i} p_i \rho^A_i \otimes \rho^B_i
\end{equation}
where $\rho^A_i$ and $\rho^B_i$ are local quantum states. However, we will instead allow the local operators appearing in equation (\ref{sepfirst}) to come from `cylindrical' state spaces, and this will change the class of states that we can consider separable. A `cylinder of radius $r$' is defined as the following set of normalised (i.e. unit trace) operators:
%\begin{widetext}
%\begin{equation}
\begin{multline}
%\hspace*{-0.7cm}
{\rm Cyl}(r) := \{ \rho |\rho=\rho^{\dag},\tr\{\rho\}=1,\\  x^2 + y^2 \leq r^2, z \in [-1,1] \}
\end{multline}
%\end{equation}
%\end{widetext}
%
where $x,y,z$ are the Bloch expansion coefficients of $\rho$, i.e. $\rho = (I + xX + yY + zZ)/2$, where $X,Y,Z$ are the Pauli operators and $I$ is identity. Visually this is a set of Bloch vectors drawn from cylinders of radius $r$, hence the name (see figure \ref{cylinderpic}).
For $r>0$ these state spaces always contain non-quantum states, as whatever the value of $r>0$, they protrude from the Bloch sphere at the poles - for the same reason they
always contain some pure qubit states.
The cylinder sets can also be rewritten in terms of a norm:
\begin{equation}
{\rm Cyl}(r) := \{ \rho |  \rho=\rho^{\dag}, \tr\{\rho\}=1, \| \rho - \mathcal{D}_Z(\rho) \| \leq r \}
\end{equation}
where $\mathcal{D}_Z(\rho)$ is the dephasing of $\rho$ (i.e. with all off-diagonal elements replaced by zero) and the norm is the trace norm. We may also define
${\rm Cyl}(r)$ in terms of a dephasing transformation on ${\rm Cyl}(1)$:
\begin{equation}
{\rm Cyl}(r) := \{ \rho| \rho = r \sigma+(1-r)\mathcal{D}_Z\left(\sigma \right), \sigma \in {\rm Cyl}(1) \}
\end{equation}
%\begin{equation}
%{\rm Cyl}(r) := r{\rm Cyl}(1) + (1-r)\mathcal{D}_Z\left({\rm Cyl}(1)\right)
%\end{equation}
(this version will be the one we will use for generalising our results to qudit systems).

Much of the intuition behind the paper can be understood from the following part of Lemma 1:

\medskip

{\bf Lemma 1 (part of):} Consider a $CZ$ gate that acts on input operators that are drawn from `cylinders' of radius $r$. The output can be
given a separable decomposition if the operators in the separable decomposition are drawn from cylinders of radius $\lambda r$, where $\lambda = \sqrt{{1 \over  \sqrt{5} - 2}} \approx 2.06$.

\medskip
We call the constant $\lambda$ (and generalisations that we later describe) a {\it disentangling growth rate}, sometimes prefixing the word `cylinder' or `cylindrical' in order to explicitly emphasise the type of entanglement we are considering.

As we shall now describe, we may combine Lemma 1 with classical algorithms that have been developed for systems with limited quantum entanglement, to obtain classical simulation algorithms for the entangled quantum systems that we consider in this work.

Let us first recap the main intuition behind why multiparticle quantum separable states might in some cases be classically easy to simulate. If a state of multiple particles (say $A, B, C, ...$) is well approximated by a quantum separable decomposition (perhaps with some grouping into blocks of particles),
\begin{equation} \label{qsep}
\rho_{ABC...} \approx \sum_{i} p_i \rho^A_i \otimes \rho^B_i \otimes \rho^C_i \otimes ...
\end{equation}
then one could attempt to sample the outcomes of local measurements on the state in by first sampling the classical distribution $p_i$ and then sampling the outcomes of local measurements on the $i$th product state (which can be efficiently described by a linear number of small matrices, as opposed to the exponentially large ones that would be required to describe arbitrary quantum states).
Although this approach seems plausible, it may fail to be efficient if the classical distribution $p_i$ cannot be efficiently sampled, or if the errors in any discrete approximations accumulate uncontrollably. In spite of this, the underlying intuition has been shown to work well in a variety of situations where the entanglement present in the system is very limited \cite{ABnoisy,Jozsa,HN}. 

%Similarly, if there is entanglement, but it is bounded or controlled in an appropriate way then efficient classical simulation may still be possible \cite{Vidal,peps}.

Of course, entangled quantum states cannot usually be well approximated by a separable decomposition of the form of (\ref{qsep}) in the first place, and so for such systems one would not expect such an approach to classical simulation to work. However, for the variants on cluster state quantum computation considered in this work, we will see that one can write down a suitable cylinder-separable decomposition upon which an efficient classical simulation can be constructed. We give an informal overview in this section, technical details and generalisations to qudits are explained in later sections.

Suppose that the input single qubit states $\ket{\psi}$ to a finite depth circuit of $CZ$s are pure qubit quantum states drawn from within a cylinder of radius $r$. Each time we apply a $CZ$ gate in our circuit, on the basis of Lemma 1 we may update the state of the particles so that if the inputs at that point have radius $r_{in}$, they are replaced by a product of two new cylinder states of radius $r_{out} = \lambda r_{in} \approx 2.06 r_{in}$, which are sampled from
the cylinder separable decomposition. Each time a qubit undergoes a $CZ$ interaction we update its state, and each time the radius grows by a factor of $\lambda$. In this way we always have a product decomposition of the system, as opposed to one involving exponentially large matrices. However this comes at a cost, the radius of the product operators in the decomposition will be
\begin{equation}
  \lambda^D r \approx (2.06)^D r
\end{equation}
where $D$ is the degree of the cluster lattice (i.e. the maximum number of edges touching a vertex), and if $D$ is large these operators will be far from physical states. However, it turns out that we may still use these non-physical states to efficiently sample the allowed measurements (i.e. $Z$ and $XY$ plane measurements) using the algorithm of \cite{HN} provided that the radius does not grow beyond 1 for any particle in the system, because {Cyl(1)} corresponds to what is referred to as the (normalised) {\it dual} of the allowed measurements. The `dual' of a set $\Omega$ of operators is the set 
\begin{equation} \label{normdual}
\Omega^*:=\{\sigma| \tr\{\sigma^{\dag} \omega \}\geq 0, \forall \omega \in \Omega \},     
\end{equation}
so in physical terms the dual of a set of measurements is the set of operators that return positive probabilities for those measurements under the Born rule - in our case we will exclusively consider normalised duals, by which we also add the constraint that the operators are unit trace. All of the foregoing discussion means that provided:
\begin{equation}
 \lambda^D r \leq 1      \,\,\,\,\, \Rightarrow \,\,\,\,\, r \leq {1 \over \lambda^D} \approx {1 \over (2.06)^D}
\end{equation}
we can efficiently simulate classically. 
%Cylinders of these radii contain pure states that are not computational basis states, and this means that the method simulates efficiently some pure entangled non-stabilizer states.
The method, and many of the intuitions underlying it, generalise to a variety of other systems, as we discuss in section \ref{section_generalisation}, and are amenable to a form of coarse graining as we discuss in section \ref{SectionCoarse}. This can significantly increase the size of the classically simulatable region for lattices with sufficient structure.
%We call such systems {\it privileged basis architectures}.
%%%%%%%%%%%%%%%%%%%%%%%%%%%%%%%%%%%%%%%%%%%%%%%%%%%%%%%%%%%%%%%%%%%%%%%%%%%%%%%
\section{Structure of the paper}

This paper is structured as follows. In section \ref{Mainlemma} we prove lemma 1. In section \ref{Classsim} we explain the classical simulation algorithm that exploits generalised separability. In section \ref{section_generalisation} we provide the generalisation of lemma 1 to various other systems, including ones in which the $CZ$ interactions are replaced by any other diagonal multi-qubit gates, and analogous systems using qudits (`privileged basis systems'). In section \ref{SectionCoarse} we explore a coarse grained version of the simulation method which increases the range of classically simulatable inputs, as well as bringing connections to the foundations of physics. In section \ref{SectionObstacles} we discuss obstacles facing attempts to classically simulate efficiently an increased range of qubit input states. We conclude in section \ref{SectionDiscussion} with a summary and discussion on the extent to which the results may be generalised further. The appendices contain some computations that we defer from the main text. Readers who only want the short version can get much of the intuition from figures 1 to 3.

%%%%%%%%%%%%%%%%%%%%%%%%%%%%%%%%%%%%%%%%%%%%%%%%%%%%%%%%%%%%%%%%%%%%%%%%%%%%%%%
\section{Maintaining a separable decomposition by increasing the radius} \label{Mainlemma}

In order to develop the classically efficient simulation, we need to achieve two things. Firstly, we need to show how a generalised separable decomposition can be obtained, and then we need to argue that it can be efficiently simulated. The latter point follows almost immediately from the classical simulation method described in \cite{HN}, albeit with some efficiencies possible due to fact that our circuits have a simpler structure. We defer discussion of this to a later section. In this section we concentrate on the first task, by establishing the main technical tool that we will need to obtain a separable decomposition in terms of cylinder state spaces.

The key observation that if a $CZ$ gate acts on two cylindrical state spaces, then the output is separable w.r.t. two new cylindrical state spaces with larger radius. This is expressed by the following lemma:

\medskip
{\bf Lemma 1: (Cylinder disentangling growth rates)} Consider the set $CZ( {\rm Cyl}(r_A) \otimes {\rm Cyl}(r_B))$ of two qubit operators made by acting with a $CZ$ gate on ${\rm Cyl}(r_A) \otimes {\rm Cyl}(r_B)$. Any operator in $CZ( {\rm Cyl}(r_A) \otimes {\rm Cyl}(r_B))$  can be written in the generalised separable form:
\begin{equation}
\sum_i p_i \rho^A_i \otimes \rho^B_i \label{csep}
\end{equation}
where $\rho^A_i \in {\rm Cyl}(R_A)$ and $\rho^B_i \in {\rm Cyl}(R_B)$ if and only if:
\begin{equation} \label{lem1}
1 \geq  \left({r_A \over R_A}+{r_B \over R_B}\right)^2 + \left({r_A \over R_A}\right)^2\left({r_B \over R_B}\right)^2
\end{equation}
We refer to an operator of the form of equation (\ref{csep}) as being ${\rm Cyl}(R_A),{\rm Cyl}(R_B)$-separable. Note that as the cylinders are the convex hulls of their extremal points, we may pick the $\rho^A_i$ and $\rho^B_i$ appearing in the decomposition (\ref{csep}) to have radii exactly equal to $R_A$ and $R_B$ respectively.

\medskip

\noindent Before we prove Lemma 1 let us discuss its interpretation.
Let us define the ratios
\begin{eqnarray}
g_i := {R_i \over r_i}
\end{eqnarray}
and refer to them as `disentangling growth rates'. Roughly speaking, the lemma states that the $CZ$ can be interpreted as a gate giving separable output,
provided that the radii of the output spaces are sufficiently large relative to those of the input spaces, i.e. provided that the disentangling growth rates are large enough. It may be helpful to note that
the constraint (\ref{lem1}) is very closely approximated by the constraint one gets by disregarding the low magnitude fourth order terms:
\begin{equation}
1 \gtrsim  {r_A \over R_A}+{r_B \over R_B}
\end{equation}
In terms of growth factors this gives:
\begin{equation}
1 \gtrsim  {1 \over g_A}+{1 \over g_B}
\end{equation}
So if $g_A$ is small, then $g_B$ must be large, and vice versa.

\begin{figure}[ht!]
\centering
%\hspace{-1cm}
%\includegraphics[width=100mm]{czgrowth.eps}
\begin{tikzpicture}[scale=.48]
%GRID AND CZ GATE
\draw[help lines, color=gray!30, dashed] (-3,0) grid (14,4);
\node at (-1.5,2) {\Large$\textbf{CZ:}$};
%SPHERE #1
%draw outline/perimeter of circle 
\shade[ball color = gray!40, opacity = 0.5] (1,2) circle (1cm);
\draw (1,2) circle (1cm);
% XY plane, starting at -2,0  
\draw (0,2) arc (180:360: 1 and 0.3);
% dashed line
\draw[dashed] (2,2) arc (0:180:1 and 0.3);
%node in middle
\fill[fill=black] (1,2) circle (1pt);
%CYLINDER
%vertical lines down
\draw[->,thick] (1,3) -- node[above]{$r$} (1.2,3);
\draw (0.8,3) -- (0.8,1);
\draw (1.2,3) -- (1.2,1);
%top cylinder: ellipse
\draw (1,3) ellipse (0.2 and 0.05);
%arc at bottom & dashed arc: starts at (0.8,1) - size  
\draw (0.8,1) arc (180:360:0.2 and 0.05);
\draw [dashed] (0.8,1) arc (180:360: 0.2 and -0.05);

%2ND PLOT
%SPHERE
%draw outline/perimeter of circle 
\shade [xshift=3cm][ball color = gray!40, opacity = 0.5] (1,2) circle (1cm);
\draw [xshift=3cm](1,2) circle (1cm);
% XY plane, starting at -2,0  
\draw [xshift=3cm](0,2) arc (180:360: 1 and 0.3);
% dashed line
\draw[xshift=3cm][dashed] (2,2) arc (0:180:1 and 0.3);
%node in middle
\fill[xshift=3cm][fill=black] (1,2) circle (1pt);
\draw [xshift=3cm] [->,thick] (1,3) -- node[above]{$r$} (1.2,3);
\draw [xshift=3cm](0.8,3) -- (0.8,1);
\draw [xshift=3cm](1.2,3) -- (1.2,1);
%top cylinder: ellipse
\draw [xshift=3cm](1,3) ellipse (0.2 and 0.05);
%arc at bottom & dashed arc: starts at (0.8,1) - size  
\draw [xshift=3cm](0.8,1) arc (180:360:0.2 and 0.05);
\draw [xshift=3cm][dashed] (0.8,1) arc (180:360: 0.2 and -0.05);

%ARROW 
\draw[->, thick] (5.5,2) -- (7.5,2);

%3RD PLOT
%SPHERE
%draw outline/perimeter of circle 
\shade [xshift=8cm][ball color = gray!40, opacity = 0.5] (1,2) circle (1cm);
\draw [xshift=8cm](1,2) circle (1cm);
% XY plane, starting at -2,0  
\draw [xshift=8cm] (0,2) arc (180:360: 1 and 0.3);
% dashed line
\draw [xshift=8cm] [dashed] (2,2) arc (0:180:1 and 0.3);
%node in middle
\fill [xshift=8cm] [fill=black] (1,2) circle (1pt);
%CYLINDER
\draw [xshift=8cm] [->,thick] (1,3) -- node[above right]{$\lambda r$} (1.4,3);
\draw [xshift=8cm](0.6,3) -- (0.6,1);
\draw [xshift=8cm](1.4,3) -- (1.4,1);
%top cylinder: ellipse
\draw [xshift=8cm](1,3) ellipse (0.4 and 0.1);
%arc at bottom & dashed arc: starts at (0.6,1) - size  
\draw [xshift=8cm](0.6,1) arc (180:360:0.4 and 0.1);
\draw [xshift=8cm][dashed] (0.6,1) arc (180:360: 0.4 and -0.1);

%4RD PLOT
%SPHERE
%draw outline/perimeter of circle 
\shade [xshift=11cm][ball color = gray!40, opacity = 0.5] (1,2) circle (1cm);
\draw [xshift=11cm](1,2) circle (1cm);
% XY plane, starting at -2,0  
\draw [xshift=11cm](0,2) arc (180:360: 1 and 0.3);
% dashed line
\draw [xshift=11cm][dashed] (2,2) arc (0:180:1 and 0.3);
%node in middle
\fill [xshift=11cm][fill=black] (1,2) circle (1pt);
%CYLINDER 
\draw [xshift=11cm] [->,thick] (1,3) -- node[above right]{$\lambda r$} (1.4,3);
\draw [xshift=11cm](0.6,3) -- (0.6,1);
\draw [xshift=11cm](1.4,3) -- (1.4,1);
%top cylinder: ellipse
\draw [xshift=11cm](1,3) ellipse (0.4 and 0.1);
%arc at bottom & dashed arc: starts at (0.6,1) - size  
\draw [xshift=11cm](0.6,1) arc (180:360:0.4 and 0.1);
\draw [xshift=11cm][dashed] (0.6,1) arc (180:360: 0.4 and -0.1);

\end{tikzpicture}

\caption{When we apply a $CZ$ (control-$Z$) operation to two input cylinders, the output can be given a separable decomposition with respect to `cylindrical' state spaces provided that the cylinder radius grows by $\lambda \approx 2.06$.
In particular the $CZ$ be described as a probabilistic transformation mapping products of input cylinder operators with radius $r$ to products of output cylinder operators of radius $\lambda r$. We call
this the `stochastic' representation. This means that if each qubit undergoes a finite number of $CZ$ gates, and if we start with inputs (which can be pure) drawn from a narrow enough radius, the overall output state can be represented as cylinder separable states with $r \leq 1$. This enables a classically efficient sampling algorithm because such operators return positive probabilities on the permitted measurements (in the $Z$ direction and $XY$ planes). The approach is amenable to a form of coarse graining and applies to all finite degree lattices consisting of (i) diagonal gates in the computational basis, (ii) local destructive measurements in the computational basis or in bases unbiased to it.}
\label{czgrowth}
\end{figure}
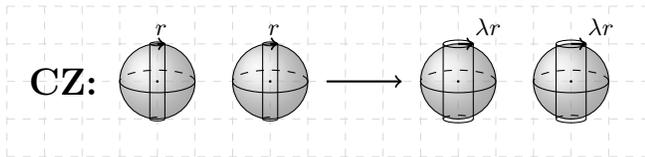

We will mostly consider the symmetric case where $g_A = g_B = g$. In this
case equation (\ref{lem1}) becomes (exactly)
\begin{eqnarray}
1 - {4 \over g^2} - {1 \over g^4} \geq 0 \nonumber
\end{eqnarray}
which can be solved to give (using the fact that $g \geq 0$ by definition anyway):
\begin{equation} \label{lambda}
g \geq \lambda := \sqrt{{1 \over  \sqrt{5} - 2}} \approx 2.05817 \nonumber
\end{equation}
So we see that as long as the radii of the output spaces are roughly twice the input radii, then the $CZ$ can be considered a separable operation (see figure \ref{czgrowth}).
%%FIGURE

\subsection{Proof of Lemma 1}

Consider a two particle operator $\rho_{AB}$. We may expand it in the Pauli basis
as
\begin{equation*}
\rho_{AB} = {1 \over 4}\sum_{i,j} \rho_{i,j} \sigma_i \otimes \sigma_j
\end{equation*}

where $\sigma_0 = I, \sigma_1 = X, \sigma_2=Y, \sigma_3= Z$ are the four Pauli matrices. Whenever expansion coefficients refer
to a Pauli operator expansion we will use square brackets ``$[$'',``$]$'', reserving curved brackets ``$($'',``$)$''
for expansion coefficients in the computational basis or for basis independent descriptions. So for instance we will display the coefficients $\rho_{i,j}$ as a 4 x 4 matrix in square brackets, with rows and columns numbered
from 0,..,3:
\begin{eqnarray*}\left[\begin{array}{cccc}
     \rho_{00}=1 & \rho_{01} & \rho_{02} & \rho_{03} \\
     \rho_{10} & \rho_{11} & \rho_{12} & \rho_{13} \\
     \rho_{20} & \rho_{21} & \rho_{22} & \rho_{23} \\
     \rho_{30} & \rho_{31} & \rho_{32} & \rho_{33}
      \end{array}\right]
\end{eqnarray*}
where we have assigned $\rho_{00}=1$ as we will consider normalised operators.
When we are considering products of local normalised operators, we will use the notation (again with square brackets):
\begin{equation*}
[1, x_A, y_A, z_A] \otimes [1, x_B, y_B, z_B]
\end{equation*}
to denote the product operator
\begin{multline*}
%\begin{equation*}
{1 \over 2} (\sigma_0 + x_A \sigma_1 +  y_A \sigma_2 +  z_A \sigma_3)\\ \otimes {1 \over 2} (\sigma_0 + x_B \sigma_1 +  y_B \sigma_2 +  z_B \sigma_3)
%\end{equation*}
\end{multline*}

Let us consider a two particle product state $\rho_A \otimes \rho_B$ where $\rho_A, \rho_B$ are drawn from two cylinders with radii
$r_A$ and $r_B$ respectively. Our goal is to determine whether the output of a $CZ$ gate, acting on
all such possible inputs, leads to a ${\rm Cyl}(R_A),{\rm Cyl}(R_B)$-separable state.

With this goal in mind we only need to consider extremal points, because if the output from all extremal
inputs is ${\rm Cyl}(R_A),{\rm Cyl}(R_B)$-separable, then the output will be separable for all inputs because the $CZ$ is linear.

Furthermore, we may exploit the symmetry about the $Z$ axis, as follows.
Suppose that we can provide an explicit ${\rm Cyl}(R_A),{\rm Cyl}(R_B)$-separable decomposition:
\begin{equation*}
CZ(\rho_A \otimes \rho_B) = \sum_{i} p_i \omega^i_A \otimes \omega^i_B
\end{equation*}
where $\omega^i_y \in {\rm Cyl}(R_y)$. Then because $CZ$ commutes with local $Z$ rotations $U_z$, and
because the cylinders are invariant under $Z$ rotations, we automatically have the ${\rm Cyl}(R_A),{\rm Cyl}(R_B)$-separable decomposition
\begin{equation*}
CZ(U^A_z(\rho_A) \otimes U^B_z(\rho_B)) = \sum_{i} p_i U^A_z(\omega_A) \otimes U^B_z(\omega_B)
\end{equation*}
By exploiting this $Z$ rotation equivalence, w.l.o.g. we may restrict our attention to input products with expansions
of the form $[1, r_{A/B},0,\pm 1]$.

We may now make one further simplification. It is easy to verify that if the first input particle has $z=1$, and output is separable:
\begin{equation*}
CZ( [1, r_A,0,1] \otimes [1, r_B,0,\pm 1] ) = \sum_{i} p_i \omega^i_A \otimes \omega^i_B
\end{equation*}
then modifying the first input to have $z=-1$ gives another operator with a separable decomposition:
\begin{multline*}
%\begin{equation*}
CZ( [1, r_A,0,-1] \otimes [1, r_B,0,\pm 1] ) \\ 
= \sum_{i} p_i  X \omega^i_A X^{\dag} \otimes Z \omega^i_B  Z^{\dag}
%\end{equation*}
\end{multline*}
In this argument we could equally well have considered the second input instead, as the $CZ$ is symmetric. This means that we need only consider one input extremum:
\begin{equation*}
[1, r_A,0,1] \otimes [1, r_B,0,1]
\end{equation*}
and determine whether the output is ${\rm Cyl}(r_A),{\rm Cyl}(r_B)$-separable. Under the action of the $CZ$ gate this input transforms to:
\begin{eqnarray} \label{output}
\left[\begin{array}{cccc}
     1 & r_B & 0 &  1 \\
     r_A & 0 & 0 &  r_A \\
     0 & 0 & r_A r_B & 0 \\
     1 &  r_B & 0 & 1
      \end{array} \right]
\end{eqnarray}
If this corresponds to a ${\rm Cyl}(R_A),{\rm Cyl}(R_B)$-separable operator, then it can be written as the outer product:
%\begin{widetext}
\begin{eqnarray*}
\hspace*{-0.7cm}
\sum_i p_i  \left[\begin{array}{c}
     1  \\
     R_A \cos(\theta_i)  \\
     R_A \sin(\theta_i)  \\
     1
      \end{array} \right] \left[\begin{array}{cccc}
     1  &  R_B \cos(\phi_i) & R_B\sin(\phi_i) & 1
      \end{array} \right]
\end{eqnarray*}
%\end{widetext}
where the angles $\theta_i$ and $\phi_i$ indicate where on the top perimeter of the cylinder the local states lie.
If we left multiply the previous two equations by
\begin{eqnarray*}
\left[\begin{array}{cccc}
     1 & 0 & 0 & 0 \\
     0 & 1/R_A & 0 & 0 \\
     0 & 0 & 1/R_A & 0 \\
     0 & 0 & 0 & 1
      \end{array} \right]
\end{eqnarray*}
and right multiply by
\begin{eqnarray*}
\left[\begin{array}{cccc}
     1 & 0 & 0 & 0 \\
     0 & 1/R_B & 0 & 0 \\
     0 & 0 & 1/R_B & 0 \\
     0 & 0 & 0 & 1
      \end{array} \right]
\end{eqnarray*}
we see that equation (\ref{output}) is ${\rm Cyl}(R_A),{\rm Cyl}(R_B)$-separable iff
\begin{eqnarray} \label{output2}
\left[\begin{array}{cccc}
     1 & {r_B \over R_B} & 0 & 1\\
     {r_A\over R_A} & 0 & 0 &   {r_A\over R_A}\\
     0 & 0 & {r_Ar_B \over R_A R_B}& 0 \\
     1&  {r_B\over R_B} & 0 & 1
      \end{array}\right]
\end{eqnarray}
is ${\rm Cyl}(1),{\rm Cyl}(1)$-separable. We will now see that determining this is equivalent to
checking the usual quantum separability of a two qubit quantum operator. This can be seen as follows.
Observe that if
\begin{eqnarray} \label{output3}
\left[\begin{array}{cccc}
     1 & \pm {r_B \over R_B} & 0 & 0\\
     \pm {r_A\over R_A} & 0 & 0 &  0 \\
     0 & 0 & {r_Ar_B \over R_A R_B}& 0 \\
     0&  0 & 0 & 0
      \end{array}\right]
\end{eqnarray}
has a quantum separable decomposition,
\begin{equation*}
\sum_i p_i [1, x^i_A,y^i_A,z^i_A] \otimes [1, x^i_B,y^i_B,z^i_B]
\end{equation*}
then (\ref{output2}) is ${\rm Cyl}(1),{\rm Cyl}(1)$-separable because it has decomposition
\begin{equation*}
\sum_i p_i [1, x^i_A,y^i_A,1] \otimes [1, x^i_B,y^i_B,1]
\end{equation*}
Moreover, by taking any ${\rm Cyl}(1),{\rm Cyl}(1)$-separable decomposition for equation (\ref{output2}) and setting
$z^i_A=0$ and $z^i_B=0$ for all $i$, we recover a quantum-separable decomposition for (\ref{output3}).
This means that (\ref{output}) is ${\rm Cyl}(r_A),{\rm Cyl}(r_B)$-separable if and only if (\ref{output3}) corresponds to a positive and PPT operator, so we may apply the PPT criterion \cite{PPT1,PPT2}.

Written out explicitly, verifying this corresponds to checking that the minimal eigenvalues of the operator given by equation (\ref{output3})
\begin{equation*}
I + ({r_A\over R_A}X \otimes I + I \otimes {r_B\over R_B}X) + {r_Ar_B \over R_A R_B} Y \otimes Y
\end{equation*}
and its partial transpose
\begin{equation*}
I + ({r_A\over R_A}X \otimes I + I \otimes {r_B\over R_B}X) - {r_Ar_B \over R_A R_B} Y \otimes Y
\end{equation*}
are non-negative. The eigenvalues of these operators can be found quite straightforwardly. We first note that these two operators can be interconverted by applying an $X$ transformation on the first qubit (as it changes $Y \otimes Y$ to $-Y \otimes Y$, but leaves the other terms alone). Hence the eigenvalues of the two operators are equivalent and so we only need to work out the eigenvalues of one equation, say the second one.
For convenience we apply a Hadamard unitary to both qubits to give
\begin{eqnarray}
I + ({r_A\over R_A}Z\otimes I + I \otimes {r_B\over R_B}Z) - {r_Ar_B \over R_A R_B} Y \otimes Y
\end{eqnarray}
Explicitly, in the computational basis, this is the matrix
\begin{widetext}
\begin{eqnarray*} \left(\begin{array}{cccc}
     1+f_A+f_B & 0 & 0 & f_Af_B \\
     0 & 1 + f_A- f_B & -f_Af_B & 0\\
     0 & -f_Af_B & 1-f_A+f_B & 0 \\
     f_Af_B & 0 & 0 & 1-f_A-f_B
      \end{array}\right), \label{xy}
\end{eqnarray*}
\end{widetext}
where we have defined $f_A:=r_A/R_A$ and $f_B:=r_B/R_B$.
The eigenvalues can be worked out on the inner and outer block separately. Both blocks have positive trace. The determinants of the inner and outer block are:
\begin{eqnarray}
1 - (f_A-f_B)^2 - f_A^2f_B^2 \nonumber \\
1 - (f_A+f_B)^2 - f_A^2f_B^2 \nonumber
\end{eqnarray}
respectively. As $f_A,f_B \geq 0$, the lowest of these is the outer determinant, so if the outer determinant is non-negative the output will be ${\rm Cyl}(R_A),{\rm Cyl}(R_B)$-separable.
This completes the proof of Lemma 1. We remark that many of the ingredients of the proof can be applied to other control-phase gates on qubits. $\blacksquare$.

It is useful to note that the separable action of the $CZ$ gate on cylinder states can be represented in quite an efficient way as a radius increase accompanied by a $z$-dependent probabilistic application of unitaries. We refer to this representation as the `{\it stochastic representation}', and describe it in an appendix as it is not essential for the majority of our discussion.
%%%%%%%%%%%%%%%%%%%%%%%%%%%%%%%%%%%%%%%%%%%%%%%%%%%%%%%%%%%%%%%%%%%%%%%%%%%%%%%
\section{Classical simulation algorithm based upon uniform disentangling growth rates} \label{Classsim}

In this section we describe the classical algorithm. We will primarily focus on the qubit case, but up to relatively minor technical adjustments the method works for all PBS systems (we discuss these technical adjustments in the next section). This is an overview of the structure of the algorithm:
\begin{enumerate}
    \item {\bf Inputs:} A desired target total variation distance $\epsilon$. The adjacency matrix describing the layout of the qubits and interactions. A classical description of the measurement pattern, which can be adaptive, as long as the measurements are drawn from the permitted set (computational basis measurements and measurements in the $XY$ plane). A classical description of the initial single particle state of each qubit (prior to the $CZ$s being applied) - the qubits must be drawn from cylinders of radius $r \leq \lambda^{-D}$. The number of bits of precision $l$ with which real parameters must be described is determined by the algorithm of \cite{HN}, and will be described shortly.
    \item {\bf Outputs:} A sample $x \in \mathbb{Z}_2^n$ from a probability distribution $p(x)$ that approximates the actual quantum distribution $q(x)$ such that total variation distance $\sum_{x} |p(x)-q(x)| < \epsilon$.
    \item {\bf Runtime:} The runtime is $O(poly(n)/\epsilon)$, inherited from the algorithm of \cite{HN}, which we leverage in this work.
\end{enumerate}

Let us first review the algorithm of \cite{HN}, upon which our approach is based. The algorithm of \cite{HN} was built upon the notion of quantum separability in a gate model framework. We will adapt it to our situations, where we use generalised separability in the MBQC framework. We will only cover the technical details of \cite{HN} that we need, for a full description see \cite{HN}.

Consider a quantum device consisting of $n$ quantum particles, initialised in a product state, undergoing a polynomial number of noisy quantum gates that do not generate any quantum entanglement, followed by local measurements of each particle. In \cite{HN} an efficient classical simulation method was proposed for such systems. Its output is a sample from a probability distribution that is within total variation distance $\epsilon$ of the probability distribution of the measurement outcomes, and it produces this output in time $O(poly(n)/\epsilon)$. We will combine the algorithm of \cite{HN} with the notion of cylinder separability to obtain a classically efficient simulation algorithm for the variants of cluster state computation that we consider in this paper. 
We remark that the term `{\it efficient}' in `{\it efficient classical simulation}' is used in a wide variety of ways in the literature (see \cite{Hakop} for a detailed discussion). Our classical algorithm inherits its  $O(poly(n)/\epsilon)$ runtime from \cite{HN} and falls under the category of an `$\epsilon$-simulation' as defined by \cite{Hakop}.

We will explain the algorithm of \cite{HN} in the context of qubits/cylinders, as the version we will need for qudit systems proceeds analogously (apart from one technical consideration which we defer to the paragraph following  equation (\ref{generalcyl})). In \cite{HN} each input qubit is represented by its Bloch vector, stored to $l$ bits of precision. We will discuss how $l$ is chosen shortly, however for now we just treat it as a parameter. Suppose w.l.o.g. that the first gate in the circuit is $\mathcal{E}$, acting upon qubits $A$ and $B$. In the actual quantum circuit this corresponds to a transformation of the form:
\begin{equation} \label{gatestep}
\mathcal{E}(\rho_A \otimes \rho_B) = \sum_i p_i \rho^i_A \otimes \rho^i_B
\end{equation}
due to the fact that $\mathcal{E}$ preserves separability. The algorithm represents this through a {\it gate simulation step} which takes as inputs $l$ bit approximations of $\rho_A$ and $\rho_B$ and constructs (through a brute force search over candidate decompositions, possible by Carath\'{e}odory's theorem) an approximation to the decomposition on the right hand side of equation (\ref{gatestep}), in which the $p_i$ and the Bloch vectors of $\rho^i_A, \rho^i_B$ are represented to $l$ bit precision. The algorithm then samples $i$ from the approximate $p_i$s, and then updates Bloch vectors of qubits $A,B$ with the $l$ bit approximations to the Bloch vectors of $\rho^i_A, \rho^i_B$ for the value of $i$ returned from that sampling. Hence the state of the system after the first gate, as represented by the algorithm, remains a product of approximate Bloch vectors for each qubit. 

The algorithm then repeats this gate simulation step for each subsequent gate, so that the state of each qubit is always stored as an $l$ bit approximation of a Bloch vector. At the end of the algorithm the measurement outcomes are sampled from the final product state using the Born rule. The analysis of \cite{HN} shows how to pick a value of $l$ of order $O(\log (poly(n)/\epsilon))$ (where $n$ is the number of gates in the circuit) such that their classical algorithm samples the quantum distribution to within $\epsilon$ while remaining polynomial time.

In terms of applying the algorithm to our setting, we note that it works for any notion of separability provided that the state space is within the normalised dual of the permitted measurements, and provided that the extremal points of the state space are given to us with $l \sim O(\log (poly(n)/\epsilon))$ bits of precision. In some cases (as with the quantum state space or the qubit cylinder state spaces) the extremal points are explicit and this is straightforward. In later sections we will consider state spaces that are provided as the solutions to inequalities, in which case the extreme points must be constructed. However under mild conditions there are well known ways to do this that we later discuss (see discussion after equation (\ref{generalcyl}))

%CAN WE USE MARGINALS IN ONE WAY OR ANOTHER TO DO A STRONG SIMULATION? APPLY THE GATES TO ONE QUBIT AND THEN COMPUTE THE P'S ON THAT QUBIT, ETC....

Let us now describe explicitly the classical simulation based upon cylinders.
Consider placing qubits at the nodes of a particular graph. Let us suppose that each qubit $i$ is initialised from within ${\rm Cyl}(r_i)$, where we will call $r_i$ the local `radius'. Now we consider applying a $CZ$ gate to two of the qubits, say qubits $1$ and $2$. The output remains cylinder separable, provided that we grow the cylinders in a way that respects equation (\ref{lem1}). Let us assume that we do this in a symmetric manner, i.e. we replace ${\rm Cyl}(r_1) \rightarrow {\rm Cyl}(\lambda r_1)$ and ${\rm Cyl}(r_2) \rightarrow {\rm Cyl}(\lambda r_2)$, using $\lambda$ as defined in equation
(\ref{lambda}). We may apply the gate simulation step of \cite{HN}, except pure qubit states in the separable decomposition are now replaced by `cylinder states' from the surface of the output cylinder, e.g. of the form $[1, \lambda r_1 \cos(\theta), \lambda r_2 \sin(\theta), \pm 1]$.
Continuing in this way we see that after all the $CZ$ gates have been applied, the output will remain cylinder separable provided that the cylinder spaces ${\rm Cyl}(r_i)$ are replaced with
\begin{eqnarray}
{\rm Cyl}(\lambda^D_i  r_i)
\end{eqnarray}
where $D_i$ is the degree of node $i$ in the graph.

Now, the cylinders are certainly not quantum state spaces. However, the cylinder space of unit radius, i.e. ${\rm Cyl}(1)$, is the dual of the measurements that are permitted. This means that provided the measurements are restricted to $Z$ measurements and measurements of the form $\cos(\theta)X + \sin(\theta) Y$, and provided that $\lambda^{d_i}  r_i \leq 1$ for all $i$, then we can use the cylinder separable description as a way to sample the measurements efficiently, as the all the required ingredients of the \cite{HN} algorithm are met, only with cylinder separability rather than quantum separability. This means that provided that the initial single qubit states satisfy
\begin{eqnarray}
\|  \rho_i - (\rho_i)_{diag} \| \leq {1 \over \lambda^{d_i}}
\end{eqnarray}
then the system can be efficiently simulated classically. In particular, if the maximum degree of any node is $D$, and if all qubits are initialised in the same state $\rho$, then we have the following theorem:

\medskip

\noindent {\bf Theorem 2:} If a quantum computation involves initialising $n$ qubits in a product state $\rho^{\otimes n}$ on the sites of a lattice, and interacting qubits joined by an edge
with $CZ$ gates, then if the single qubit states $\rho$ satisfy:
\begin{eqnarray}
\|  \rho - \rho_{diag} \| \leq {1 \over \lambda^D} \,\,\,\,\,\, \lambda := \sqrt{{1 \over  \sqrt{5} - 2}} \approx 2.05817 \nonumber
\end{eqnarray}
where $D$ is the maximum degree of any node, then measurements in the $Z$ basis and the $X-Y$ plane can be sampled classically to within
additive error $\epsilon$ in $O(poly(n,{1 \over \epsilon}))$ time.

\medskip

\noindent It is clear that this bound could be improved by using asymmetric growth factors if there is suitable further structure in the interaction graph. For instance, if a qubit has a larger degree than the qubits it is joined to, then with each $CZ$ one could apply a larger growth factor to the lower degree qubits and a lower growth factor to the higher degree qubits. Another way to exploit further structure in the graph is to use the idea of coarse graining from physics - we will explore this in a later section in the context of the 2D square lattice. We also remark that it would be possible to make significant efficiency savings by exploiting the fact that the two particle gates are always $CZ$s, so we could precompute the stochastic representation (described in Appendix A) of the $CZ$ gate to required accuracy once and then apply it repeatedly (as opposed to finding a decomposition for a two particle state after each gate has been applied). However, if one wishes to consider gates that may vary (a scenario we generalise to in the next section) then this would not be possible.

The algorithm that we have proposed does not just simulate a type of quantum device, it also simulates some hypothetical devices that we will refer to as {\it cylindrical computers}, which act upon {\it cylindrical bits}: we define a cylindrical computer to be a device that places operators prepared in the extremal points of ${\rm Cyl}(r)$, for some $r>0$, at the vertices of a given lattice, then interacts them with $CZ$ gates, and measures in the $Z$ basis and $XY$ plane measurements. We discuss some properties of such `cylindrical computers' in section \ref{SectionCoarse}.

\section{Generalisation of Lemma 1 to multi-particle gates diagonal in a computational basis} \label{section_generalisation}

This section can be skipped by readers only interested in the coarse graining discussion of the next section, which can be mostly understood from the earlier discussion of qubits.

The main ingredient of our discussion has been the observation that the output state from each $CZ$ gate is separable if we consider a cylindrical state space whose radius increases
when the gate acts. In this section we show that this observation generalises to other systems that obey four conditions $(\alpha)-(\delta)$ described below.
We will then show that these four assumptions are satisfied by system that we call {\bf privileged basis system} (PBS), defined as follows:

\medskip

\noindent {\bf Definition: Privileged Basis System} (PBS): A privileged basis system is a quantum circuit with the following properties. It consists of unitary gates (which may act on more than one particle) that are diagonal in a computational basis. Each particle undergoes only a finite number of gates. After the gates have been applied, destructive measurements are performed consisting of POVM elements of one of the following forms:
\begin{enumerate}
    \item {\bf $Z$ measurement operators:} POVM elements proportional to rank-1 projectors in the computational basis $\{\ket{j}\}$
    \item {\bf Equatorial operators:} POVM elements proportional to rank-1 projectors in any basis `unbiased' to the computational basis (i.e. projectors $\{P\}$ such that $\bra{j}P\ket{j} = 1/d$, where $d$ is the qudit dimension)
\end{enumerate}
If a measurement is in the computational basis then we will call it a {\bf $Z$ measurement}, if a measurement consists entirely of equatorial POVM elements then we will call it an {\bf `equatorial measurement'}. 

For such PBS systems there is an analogue of the `cylinder' of inputs that can be efficiently simulated classically, and on the basis of this one can write down many pure entangled systems that can be efficiently simulated classically. 

The interest in these sorts of systems stems from the variety of MBQC schemes that fall into this class. In addition to the original cluster state scheme, there are others that involve different diagonal gates and $Z$/equatorial measurements, or some subset of them - examples include weighted graph states \cite{gross2007measurement}, states built from more general control phase gates \cite{KissingerW}, states built from CCZ gates \cite{MillerM1,MillerM2, Tomo}, and generalisations of the original cluster scheme to qudits \cite{Hall}. In all such systems there will be a similar `transition' as happens for the cluster systems considered in previous sections - at one extreme quantum computation is possible, but with particles initialised in an appropriate `cylinder' one can efficiently simulate classically.

\medskip

\noindent {\bf Conditions} $(\alpha)-(\delta)$ {\bf :} Suppose that we have several quantum particles undergoing a quantum gate $\mathcal{V}$ (where $\mathcal{V}$ is the superoperator corresponding to an underlying unitary matrix, i.e. $\mathcal{V}(\rho) = V\rho V^{\dag}$ for some unitary matrix $V$). Let us assume that associated to each particle $j \in 1,..,N$ there is an abelian group $G_j$ of unitaries.
We will make four assumptions about this setup. The first three $(\alpha-\gamma)$ are as follows, the fourth $(\delta)$ will be explained shortly:
\begin{enumerate}
\item[($\alpha$)] Each group $G_j$ can be averaged over a Haar measure. Denote the resultant quantum operation (i.e. applying a unitary $U_j$ drawn randomly from the Haar measure) by $\mathcal{D}_j$, which can be considered to be a kind of dephasing operation:
\begin{equation*}
    \mathcal{D}_j(\sigma) := \int U_j \sigma U^{\dag}_j dU_j
\end{equation*}
\item[($\beta$)] There is a set $\mathcal{M}_j$ of permitted POVM elements on particle $j$ which is invariant under $G_j$,
i.e. $U\mathcal{M}_jU^{\dag}  = \mathcal{M}_j$ for any $U \in G_j$. Here by `set of permitted POVM elements' we mean that any allowed complete measurement on a particle is formed from members of $\mathcal{M}_j$. For technical reasons we will additionally assume that the normalised dual of this set of measurements is bounded (see discussion below equation (\ref{generalcyl})). We do not need to distinguish between a given POVM element $M$ and $\nu M$ for any $\nu > 0$, so in fact each $\mathcal{M}_j$ can be considered a cone. We consider only destructive measurements - i.e. each particle is discarded after measurement. 

\item[($\gamma$)] The multi-particle gate $\mathcal{V}$ commutes with any product of unitaries drawn from $\bigotimes_j G_j$.
\end{enumerate}
On the basis of these assumptions, we will make the following definitions:

\begin{enumerate}
\item {\bf Phasing.} We will define a local linear operation, parameterised by a real parameter $r \geq 0$, that is a linear combination of the identity operation $\mathcal{I}$ (leaving inputs alone) and $\mathcal{D}_j$:
\begin{equation}
\mathcal{T}_j(r) := r\mathcal{I} + (1-r) \mathcal{D}_j
\end{equation}
As a transformation on input operator $\rho$ the operation $\mathcal{T}_j(r)$ acts as:
\begin{equation}
\mathcal{T}_j(r) : \rho \rightarrow r \rho + (1-r)\mathcal{D}_j(\rho)
\end{equation}
Note that this only gives a physical quantum operation when
$r \in [0,1]$, in which case it represents dephasing noise.
However, it is convenient for us to allow all non-negative $r$. When $1 \geq r > 0$ we will say that the operation
is a noisy dephasing operation, but for more general $r$ we will refer to it as a `phasing' operation.
The definition allows us to invert $\mathcal{T}_j(r)$ for $r > 0 $ with another phasing operation:
\begin{equation*}
(\mathcal{T}_j(r))^{-1} = \mathcal{T}_j(r^{-1})
\end{equation*}
and express the product of two phasing operators as:
\begin{equation} \label{composition}
\mathcal{T}_j(rs) = \mathcal{T}_j(r)\mathcal{T}_j(s)
\end{equation}

\item {\bf Local `cylinder'.} For each particle $j$ we will consider the normalised dual of the measurements, which we call a `cylinder of radius 1', defined as follows:
\begin{equation} \label{nomdualdef}
\dua_j(1) := \{ \rho |   \tr\{M \rho\} \geq 0 \forall M \in \mathcal{M}_j, \,\,\tr{\rho}=1   \}
\end{equation}
A `cylinder' of arbitrary non-negative radius $r \geq 0$ will then be defined in terms of the action of $\mathcal{T}_j(r)$ on $\cyl{1}$:
\begin{equation} \label{generalcyl}
\dua_j(r) := \mathcal{T}_j(r)(\dua_j(1))
\end{equation}

There is one subtlety that needs to be addressed with regards to using these cylinders for classical simulation. In order to apply the brute force search over candidate separable decompositions in the gate simulation step of \cite{HN} the algorithm needs the state space to be described not as the dual of a set of measurements as we have done in equation (\ref{generalcyl}), but as the convex hull of a set of extremal points, in order to exploit Carath\'{e}odory's theorem. However, we can fix this using standard methods, as follows. Given a discretisation of the allowed measurements (determined by what precision the experimenter can set their measurement device to), each permitted measurement operator provides a bounding hyperplane of the dual, and from standard considerations \cite{Avis} the extrema are the intersections of $m$ of these planes, where $m$ is the (real) dimension of the dual space. For a set of $m$ planes whose intersection defines an extremal point, we can compute the extremal point to $l$ bits of accuracy by solving the relevant linear equations provided that the allowed measurements are described to $O(l)$ bits of precision (as a finite number of arithmetic operations are needed, and we are assuming that the cylinder is bounded). Hence there will at most be $O((\exp(O(l)))^m)$ extrema to $l \sim O(\log (poly(n)/\epsilon))$ bits precision, hence giving an overall additional cost of $O((poly(n)/\epsilon))^m) \sim O(poly(n)/\epsilon)$ per cylinder. As there are at most $n$ particles in the system, and hence at most $n$ cylinders, this remains polynomial.
\end{enumerate}

One of the consequences of the above assumptions is that when $r \in [0,1]$ (in which case the phasing to radius $r$ is a conventional noisy dephasing operation) the invariance
of the permitted measurements $\mathcal{M}_j$ implies that for any $j$;
\begin{equation*}
\mathcal{T}_j(r) (\dua_j(1)) \subseteq \dua_j(1)
\end{equation*}
and hence for any $r_2 \leq r_1$:
\begin{eqnarray*}
\dua_j(r_2) = \mathcal{T}_j(r_1) \mathcal{T}_j(r_2/r_1) (\dua_j(1)) \\
 \subseteq \mathcal{T}_j(r_1) (\dua_j(1)) = \dua_j(r_1)
\end{eqnarray*}
So cylinders of a given radius contain all cylinders of smaller radius, and in particular $ \dua_j(0)$ is contained in all other cylinders at site $j$. We now make one further assumption about the gate $\mathcal{V}$:

\begin{enumerate}
\item[($\delta$)] We assume that if $\mathcal{V}$ acts on products of inputs from $\bigotimes_j \dua_j(1)$ then there is a constant $1 \geq \mu > 0$ such that
\begin{equation*}
\hspace*{-0.7cm}
\left( \bigotimes_j \mathcal{T}_j(\mu) \right) \mathcal{V} \left(\bigotimes_j \dua_j(1) \right)  \in  {\rm Sep}\left(\bigotimes_j \dua_j(1)\right)
\end{equation*}
This assumption asserts that for a given $\mathcal{V}$ there is some amount of local dephasing noise acting at each site, other than the maximal dephasing $\mathcal{T}_j(0) = \mathcal{D}_j$ itself, which makes $\mathcal{V}$ into an operation that preserves unit cylinder separability of the inputs. Let $c$ be the maximum of all $\mu$ with this property.

We can restate the assumption by acting on both sides of the equation with the inverse phasing operation to give:
\begin{equation*}
 \mathcal{V} \left(\bigotimes_j \dua_j(1) \right)  \in  {\rm Sep}\left(\bigotimes_j \dua_j({1 \over c})\right)
\end{equation*}
\end{enumerate}
The assumptions $(\alpha)-(\delta)$ have the consequence that $\mathcal{V}$ acting upon a collection of cylinders will lead to a separable output provided that the cylinder radii
are grown by a factor $1/c$, as may be seen by the following equation:
\begin{eqnarray*}
\mathcal{V} \left(\bigotimes_j \dua_j(r_j) \right) = \bigotimes_j \mathcal{T}_j(r_j) \mathcal{V} \left(\bigotimes_j \dua_j(1) \right)\\
\subseteq \bigotimes_j \mathcal{T}_j(r_j)   {\rm Sep}\left(\bigotimes_j \dua_j({1 \over c})\right)  = {\rm Sep}\left(\bigotimes_j \dua_j({r_j \over c})\right)
\end{eqnarray*}
This means that for any system obeying the assumptions $(\alpha)-(\delta)$, 
\begin{equation}
 {1 \over c}
\end{equation}
will serve as the analogue of the growth factor $\lambda$ as used in the qubit case, and will allow for an analogous classical simulation algorithm to be formulated.

\medskip

\noindent {\bf Proof (Privileged Basis Systems satisfy the conditions):}
We now show that PBS satisfy the conditions $(\alpha) - (\delta)$ . Apart from the requirement that the $\dua_j({1})$ are bounded (which is part of condition ($\beta$)), conditions $(\alpha) - (\gamma)$ are immediately satisfied if we pick each group $G_j$ to consist of the unitaries on particle $j$ that are diagonal in the computational basis. So we need to show (i) that the $\dua_j({1})$ are bounded in order to fully satisfy the $(\beta)$ condition, (ii) assumption $(\delta)$ is satisfied if we dephase using these groups.

For simplicity in our discussion we assume that all the qudits have the same dimension $d$ and use $\dua(1)$ to refer to the unit cylinder for any one particle and allowed measurements. The argument can easily be extended to situations in which the particles have different dimensions. We denote the qudit computational basis by $\ket{0},\ket{1},...,\ket{d-1}$.

Let us first explain the boundedness. In fact we will explain that any $\rho \in \cyl{1}$ must be both Hermitian and bounded. The diagonal elements of $\rho$ must be bounded as they are valid probabilities for outcomes of a $Z$ measurement. So we need only consider off-diagonal elements. W.l.o.g. we consider the element $\bra{0}\rho \ket{1}$ and write it as $\bra{0}\rho \ket{1} = t \exp(i \omega)$. The argument easily extends to other off-diagonal elements. The approach we take is to express the off-diagonal elements in terms of probabilities of measurement outcomes, and this will allow us to show both boundedness and hermiticity. Using the fact that $\rho$ is of unit trace (its diagonal forms a probability distribution) it can be verified that:
\begin{equation} \label{bounded}
    \bra{0}\rho \ket{1} + \bra{1}\rho \ket{0} = - 1 + {d \over 2^{d-2}} \sum  \bra{v} \rho \ket{v}
\end{equation}
where the sum ranges over all vectors $\ket{v}$ of the form $(\ket{0}+\ket{1} \pm \ket{2} \pm \ket{3} \pm ... \pm \ket{d-1})/\sqrt{d}$. Similarly it can be verified that:
\begin{equation} \label{bounded2}
    2t = - 1 + {d \over 2^{d-2}} \sum  \bra{\tilde{v}} \rho \ket{\tilde{v}}
\end{equation}
where the sum ranges over all vectors $\ket{\tilde{v}}$ of the form $(\exp(i \omega) \ket{0}+\ket{1} \pm \ket{2} \pm \ket{3} \pm ... \pm \ket{d-1})/\sqrt{d}$. As the vectors $\ket{v},\ket{\tilde{v}}$ are all unbiased w.r.t. to computational basis, the $\ket{v}\bra{v},\ket{\tilde{v}}\bra{\tilde{v}}$ give permitted equatorial measurement operators. Hence the rightmost sums of both equations (\ref{bounded}) and (\ref{bounded2}) are sums of probabilities, and so the right sides of (\ref{bounded}) and (\ref{bounded2}) are both real and bounded. Hence (\ref{bounded}) shows that $\rho$ is Hermitian, and (\ref{bounded2}) shows that it is bounded. Condition ($\beta$) is hence established.

Let us now turn to condition ($\delta$). To explain the argument it will be helpful to partially characterise the extremal points of the normalised dual of the allowed measurements on any one of the particles. If an operator $\rho$ is in $\dua(1)$, then its diagonal in the computational basis must be a probability distribution, as it must be in the dual of the $Z$ measurement. Now consider forming an operator $\rho'$ by replacing the diagonal of $\rho$ with another probability distribution. It is easy to see that $\rho'$ will also be in $\cyl{1}$, as it returns a valid probability distribution for $Z$ measurements, and for any of the equatorial measurements changing the probability distribution on the diagonal has no effect. This means that the extreme points of $\cyl{1}$ must have deterministic distributions on the diagonal consisting of one `1' and the rest of zeros, i.e. their diagonals must be one of $(1,0,0,...), (0,1,0,0,...)$, etc. Hence an extremal point of a single state space $\cyl{1}$ must be of the form:
\begin{equation*}
    \ket{a}\bra{a} + \sum_{m \neq n} c_{m,n} \ket {m}\bra{n}
\end{equation*}
where $\ket{a}\bra{a}$ is a computational basis state.

Now consider an $N$ qudit gate $\mathcal{V}$ that is formed from a unitary matrix $V$ that is diagonal in the computational basis. Consider acting this gate upon $N$ input qudits prepared in a product of such extrema, followed by independent noisy dephasing $\mathcal{T}_{\eta} = \eta \mathcal{I}+ (1-\eta)\mathcal{D}_G$ on each particle, where $\eta >0$ will be a small parameter whose value we will choose shortly. Our goal is to argue that if $\eta$ is small enough, the result of such an operation will be $\bigotimes^N_{j=1}\cyl{1}$-separable. 

A product of input extrema drawn from $\bigotimes^N_{j=1}\cyl{1}$ will be of the form:
\begin{equation} \label{Vout}
\ket{\underline{a}}\bra{\underline{a}} + \sum_{\underline{m}\neq \underline{n}} c_{\underline{m},\underline{n}} \ket {\underline{m}}\bra{\underline{n}}
\end{equation}
for some coefficients $c_{\underline{m},\underline{n}}$ where $\underline{a},\underline{m},\underline{n} \in \mathbb{Z}^N_d$. Note that there are other constraints on the $\underline{m},\underline{n}$ appearing in this sum, as when the first particle is (say) prepared in an extremum of the form {$\ket{0}\bra{0} +$ {\it off diag.}}, no term will appear in the sum of equation (\ref{Vout}) in which both  $\underline{m},\underline{n}$ have the same value in the first position, unless that value is $0$. Additionally, as the sum must be Hermitian we have $c_{\underline{m},\underline{n}}=c^*_{\underline{n},\underline{m}}$.

If we now act on equation (\ref{Vout}) with $V$, as $V$ is diagonal in the computational basis the form of the equation will not change. So the output of $V$ acting upon input extrema will again be of the form (\ref{Vout}), and moreover by the foregoing discussion it will be Hermitian and bounded. Now if we apply $\bigotimes^N_{j=1} \mathcal{T}_{\eta}$ to an operator of the form of equation (\ref{Vout}) off-diagonal terms will be multiplied by powers of $\eta$. Hence after the dephasing operation the state will be of the form:
\begin{equation} \label{Tout}
\ket{\underline{a}}\bra{\underline{a}} + \eta \sum_{\underline{m}\neq \underline{n}} c_{\underline{m},\underline{n}} \ket {\underline{m}}\bra{\underline{n}}
\end{equation}
where any further powers of $\eta$ have been absorbed into the coefficients $c_{\underline{m},\underline{n}}$. We will now argue that by making $\eta$ small we will force the state of (\ref{Tout}) to become generalised separable.

Let the number of ways of picking {\bf ordered} pairs $(\underline{m},\underline{n})$ such that $\underline{m} \neq \underline{n}$ be $2W$, where $W$ is a positive integer (the number of ways must be even as we can interchange $\underline{m}$ and $\underline{n}$ for any suitable choice). Consider one specific way of picking $\underline{m} \neq \underline{n}$, say $\underline{m}= \underline{x},\underline{n} = \underline{y}$ and for the corresponding {\bf unordered} pair $\{\underline{x},\underline{y}\}$ define
\begin{equation*}
    A_{\{\underline{x},\underline{y}\}} := \ket{\underline{a}}\bra{\underline{a}} + \eta  W c_{\underline{x},\underline{y}} \ket{\underline{x}}\bra{\underline{y}}+\eta W c_{\underline{y},\underline{x}} \ket {\underline{y}}\bra{\underline{x}}
\end{equation*}
Then equation (\ref{Tout}) can be rewritten as:
\begin{equation*}
   {1 \over W}  \sum_{\substack{\{\underline{x},\underline{y}\}\\ \underline{x} \neq \underline{y}}} A_{\{\underline{x},\underline{y}\}} 
\end{equation*}
where the sum is over unordered pairs $\{\underline{x},\underline{y}\}$ such that $\underline{x} \neq \underline{y}$. We will supply a separable decomposition for this expression (for sufficiently small $\eta$) by providing a separable decomposition for each $A_{\{\underline{x},\underline{y}\}}$. With this aim, let us consider one specific $A_{\{\underline{x},\underline{y}\}}$ and (to keep our equations less cluttered) rewrite it in the form
\begin{equation} \label{Eform}
    A_{\{\underline{x},\underline{y}\}} = \ket{\underline{a}}\bra{\underline{a}} + W\bigotimes_j E_j + W\bigotimes_j E^{\dag}_j
\end{equation}
where the product is over the $N$ qudits and $E_j := (\eta c_{\underline{y},\underline{x}})^{1/N} \ket{x_j}\bra{y_j}$.

The separable decomposition for equation (\ref{Eform}) will be made with a convex combination of product operators of the form:
\begin{equation} \label{sepdec}
\bigotimes_j \left( \ket{a_j}\bra{a_j}+ W^{1/N} e^{{2\pi i \over 8}v_j } E_j + W^{1/N}  e^{-{2\pi i \over 8}v_j } E^{\dag}_j \right)
\end{equation}
with real $v_j$. There are two things that we need to demonstrate: that (i) an appropriate mixture of these products gives a decomposition of (\ref{Eform}), and (ii) the local operators in each product are contained within $\cyl{1}$. The second of these points is straightforward: consider a permitted measurement operator $M$ measured on one of the factors in equation (\ref{sepdec}), the probability will be of the form:
\begin{equation*} 
\hspace*{-0.3cm}
\tr \{ M \left(\ket{a_j}\bra{a_j}+ W^{1/N} e^{{2\pi i \over 8}v_j } E_j + W^{1/N}  e^{-{2\pi i \over 8}v_j } E^{\dag}_j \right) \}
\end{equation*}
where (by the discussion following equation (\ref{Vout})) the $E_j,E_j^{\dag}$s are either off diagonal or proportional to $\ket{a_j}\bra{a_j}$. The value of this probability will be zero if $M$ is a $Z$ measurement operator not equal to $\ket{a_j}\bra{a_j}$. For all other permitted measurement operators - i.e. either $\ket{a_j}\bra{a_j}$ or the equatorial ones - the value of $\tr\{M \ket{a_j}\bra{a_j}\}$ will be $\geq 1/d$. As this is strictly positive, by making  $\eta$ small enough the potentially negative contribution from the $E_j,E_j^{\dag}$s will not make the overall probability negative. Hence for some $\eta > 0$ the local operators in equation (\ref{sepdec}) are guaranteed to be from $\cyl{1}$.

Let us now turn to point (i), checking that we can decompose the $A_{\{\underline{x},\underline{y}\}}$ as a separable mixture. Consider picking $N-1$ integers $v_j$ for $j=1,..,N-1$ from $\mathbb{Z}_8 = \{0,1,2,..,7\}$ completely at random, and then setting an $N$th integer $v_N$ to be:
\begin{equation} \label{veeN}
v_N = - \sum_{i=1}^{N-1} v_i
\end{equation}
It is not difficult to verify (as we will shortly do) that the uniform mixture of (\ref{sepdec}) over all such choices of $v_j$ equals $A_{\{\underline{x},\underline{y}\}}$:
\begin{widetext}
\begin{eqnarray*} 
    A_{\{\underline{x},\underline{y}\}} = \sum {1 \over 8^{N-1}}   \left(\bigotimes_j \left( \ket{a_j}\bra{a_j}+ W^{1/N} e^{{2\pi i \over 8}v_j } E_j + W^{1/N}  e^{-{2\pi i \over 8}v_j } E^{\dag}_j \right) \right)
\end{eqnarray*}
\end{widetext}
This expression is hence our desired generalised separable decomposition for the $A$s, and hence supplies a generalised separable decomposition for equation (\ref{Tout}) provided that $\eta$ is small enough.
For convenience we now explain how this identity arises. In order to match (\ref{Eform}), when we sum over the product operators in equation (\ref{sepdec}), we will need to eliminate `cross terms' such as:
\begin{equation}
E_1 \otimes \ket{a_2}\bra{a_2} \otimes E^{\dag}_3 \otimes ....
\end{equation}
which do not appear in (\ref{Eform}) because these `cross terms' contain basis states on some sites, $E_j$'s on other sites, and $E_j^{\dag}$s on yet others. In equation (\ref{Eform}) there are only the `non-cross' terms
\begin{equation}
\bigotimes_j \ket{a_j}\bra{a_j} \,\,\,\, , \,\,\,\, \bigotimes_j E_j \,\,\,\, , \,\,\,\, \bigotimes_j E^{\dag}_j
\end{equation}
We have picked our ensemble of $v_1,...v_8$ such that any `cross terms' cancel out, leaving only the
`non-cross terms' that appear in equation (\ref{Eform}). We can see this by substituting for $v_N$ in our separable decomposition using expression (\ref{veeN}). We find that every cross term in the separable decomposition carries a non-trivial phase related to at least one of the $v_j$s, as follows:
\begin{enumerate}
\item Any cross term containing an $\ket{a_j}\bra{a_j}$ for $j \neq N$ will contain a phase $e^{\pm {2\pi i \over 8}v_j }$ due to the contribution from $v_N$.
\item Any cross term containing $\ket{a_N}\bra{a_N}$ will contain a phase $e^{\pm {2\pi i \over 8}v_j }$ for some $j \neq N$.
\item Any other cross terms (those containing no $\ket{a_j}\bra{a_j}$ at all, only $E_j$ or $E^{\dag}_j$ contributions, but at least one of each) will have an overall phase of the form:
\begin{eqnarray*}
 \exp \left( {2\pi i \over 8}\left(\sum_{i=1}^{N-1} \pm v_i \pm \sum_{i=1}^{N-1} v_i \right) \right)
\end{eqnarray*}
such that the overall phase does not cancel, which means that for some $j \neq N$ there will be a phase contribution of
\begin{eqnarray*}
 \exp \left( \pm {2\pi i \over 8} 2v_j \right)
\end{eqnarray*}
\end{enumerate}
Altogether this means that all the cross terms contain, for some value of $j$, either a non-trivial phase contribution of the form:
\begin{eqnarray*}
 \exp \left( \pm {2\pi i \over 8} v_j \right)
\end{eqnarray*}
or one of the form
\begin{eqnarray*}
 \exp \left( \pm {2\pi i \over 8} 2v_j \right)
\end{eqnarray*}
We can exploit this to eliminate the cross terms.
In choosing the $v_j$ in the way that we have, we note when summing the phases over them we get:
\begin{eqnarray*}
\sum_{v_j} \exp \left( \pm {2\pi i \over 8} v_j \right) = \sum_{v_j} \exp \left( \pm {2\pi i \over 8} 2v_j \right) = 0
\end{eqnarray*}
Hence the cross terms cancel to leave exactly the right side of (\ref{Eform}), as desired.

All of this means that we have the following result:

\medskip

\noindent {\bf Theorem 3:} Consider {\it privileged basis systems}, i.e. suppose that we have a computational basis (the `$Z$ basis'), a set of permitted destructive measurements $\mathcal{M}$ consisting of measurements in the computational basis (`$Z$ measurements') and measurements consisting
of all unbiased rank-1 projectors (`equatorial measurements'), and suppose moreover that each qudit undergoes at most $D$ gates, which all are diagonal in the computational basis.
Then there is a $1 \geq c > 0$ such that for qudits initialised from the set
\begin{equation}
\cyl{c^D}  = (c^D \mathcal{I} + (1-c^D) \mathcal{D}_G)\cyl{1}
\end{equation}
adaptive permitted measurements can be efficiently sampled classically, and moreover has a local hidden variable model. In this equation $\cyl{1}$ represents the normalised dual of the permitted measurements,
and $\mathcal{D}_G$ represents the total dephasing operation.
Note that although we have presented the argument with the same dimension qudit at each site and the same gate at each edge, this is not necessary for the argument. $\blacksquare$

We note that the cylinder separable states in this more general setting can include pure multiparticle quantum entangled states. To see this, first note that
$\cyl{r}$ will contain pure quantum states that are superpositions in the computational basis with one dominant element, such as:
\begin{equation*}
\ket{\psi} = \sqrt{1- (d-1)\epsilon^2} \ket{0} + \epsilon \sum_{j=1}^{d-1} \ket{j}
\end{equation*}
because if $\epsilon$ is small enough then $\mathcal{T}_{1/r} (\ket{\psi}\bra{\psi})$ will be positive for the permitted measurements for similar reasons as discussed previously - the dominant
contribution from $\ket{0}\bra{0}$ will outweigh any negative contribution from any off diagonal terms. It not difficult to then construct multi-particle unitaries that are diagonal
in the computational basis that, acting upon such inputs, that will lead to pure entangled quantum states just as in the case of the $CZ$ gate.
%%%%%%%%%%%%%%%%%%%%%%%%%%%%%%%%%%%%%%%%%%%%%%%%%%%%%%%%%%%%%%%%%%%%%%%%%%%%%%%
\section{Coarse Graining} \label{SectionCoarse}

In our approach to classical simulation thus far, we faced two conflicting requirements. In order to maintain a separable decomposition, the radii must grow with each application of a $CZ$ gate. However, we have a limit to how far the radii can be grown, because they must satisfy the dual constraint, i.e. the radii must not exceed $1$ in order to not leave the dual of the permitted measurements. In this section we will see that we can increase the range of systems that can be efficiently simulated classically by managing this tradeoff better through a coarse grained approach.

It is helpful to consider the classical simulation approach we have taken more generally. Consider a set of particles undergoing a quantum circuit, and imagine that particle $i$ is initialised in a quantum state drawn from a set of operators $S_i$. Each time a gate acts, we attempt to update the state spaces to maintain a separable decomposition (this will usually involve `growing' the state spaces in some way, just as we grew the radii in the previous sections). Eventually, at the end of the circuit, we hope that the final state spaces are small enough to be inside the dual of whatever measurements we are permitting. If all steps of the scheme can be accomplished, then it could be a route to an efficient classical simulation provided that the technical requirements of \cite{HN} are also met.

Of course, given that quantum separability is a hard problem, one might expect that it will usually be too difficult to pursue this approach. However, we have a few advantages in our favour: firstly, we may try to pick state spaces for which showing separability is easy, secondly, by considering finite degree cluster-like computations each particle only undergoes a constant number of quantum-entangling gates, thirdly, we may exploit the structure of the interaction graph.
In this section we will see how the technique of coarse graining may help us exploit these advantages to improve the bounds presented above.

While the argument can be applied to any lattice for which a regular tiling of increasing size is possible, for simplicity of explanation we will consider a 2-dimensional lattice of qubits of size $N \times M$. We begin by forming the qubits into identical rectangular blocks of qubits that we will treat as single particles (we'll assume that $N$ and $M$ are chosen such that this is possible, e.g. they aren't prime). We partition the $CZ$ gates into two types: the ones acting internally within each block, and the ones acting externally that connect different blocks. We call these gates ``internal'' and ``external'' $CZ$s respectively. Now imagine that we have initialised the qubits and are about to embark upon performing the $CZ$ gates. We will analyse situations like this in a few steps, exploiting the fact that the $CZ$ gates commute, and therefore can be implemented in any order.
\begin{enumerate}
  \item We pick a starting state space for each block that is simply the product of individual cylinders ${\rm Cyl}(r)$ on each qubit.
  In a $8 \times 8$ lattice, for instance, we would have the following layout of qubits, where the dots represent our initial ${\rm Cyl}(r)$ state spaces:
  \begin{center}
  \begin{tabular}{|cccccccc|}
    \hline
  $\cdot$ & $\cdot$ & $\cdot$ & $\cdot$ & $\cdot$ & $\cdot$ & $\cdot$ & $\cdot$  \\
 $\cdot$ & $\cdot$ & $\cdot$ & $\cdot$ & $\cdot$ & $\cdot$ & $\cdot$ & $\cdot$   \\
$\cdot$ & $\cdot$ & $\cdot$ & $\cdot$ & $\cdot$ & $\cdot$ & $\cdot$ & $\cdot$  \\
 $\cdot$ & $\cdot$ & $\cdot$ & $\cdot$ & $\cdot$ & $\cdot$ & $\cdot$ & $\cdot$   \\
   $\cdot$ & $\cdot$ & $\cdot$ & $\cdot$ & $\cdot$ & $\cdot$ & $\cdot$ & $\cdot$  \\
 $\cdot$ & $\cdot$ & $\cdot$ & $\cdot$ & $\cdot$ & $\cdot$ & $\cdot$ & $\cdot$   \\
$\cdot$ & $\cdot$ & $\cdot$ & $\cdot$ & $\cdot$ & $\cdot$ & $\cdot$ & $\cdot$  \\
 $\cdot$ & $\cdot$ & $\cdot$ & $\cdot$ & $\cdot$ & $\cdot$ & $\cdot$ & $\cdot$   \\
    \hline
  \end{tabular}
  \end{center}
  \item We now partition the qubits into blocks of fixed size. For instance, we may partition the $8 \times 8$ lattice we are considering into four $4 \times 4$ blocks:
   \begin{center}
  \begin{tabular}{|cccc|cccc|}
    \hline
  $\cdot$ & $\cdot$ & $\cdot$ & $\cdot$ & $\cdot$ & $\cdot$ & $\cdot$ & $\cdot$  \\
 $\cdot$ & $\cdot$ & $\cdot$ & $\cdot$ & $\cdot$ & $\cdot$ & $\cdot$ & $\cdot$   \\
$\cdot$ & $\cdot$ & $\cdot$ & $\cdot$ & $\cdot$ & $\cdot$ & $\cdot$ & $\cdot$  \\
 $\cdot$ & $\cdot$ & $\cdot$ & $\cdot$ & $\cdot$ & $\cdot$ & $\cdot$ & $\cdot$   \\
    \hline
      $\cdot$ & $\cdot$ & $\cdot$ & $\cdot$ & $\cdot$ & $\cdot$ & $\cdot$ & $\cdot$  \\
 $\cdot$ & $\cdot$ & $\cdot$ & $\cdot$ & $\cdot$ & $\cdot$ & $\cdot$ & $\cdot$   \\
$\cdot$ & $\cdot$ & $\cdot$ & $\cdot$ & $\cdot$ & $\cdot$ & $\cdot$ & $\cdot$  \\
 $\cdot$ & $\cdot$ & $\cdot$ & $\cdot$ & $\cdot$ & $\cdot$ & $\cdot$ & $\cdot$   \\
    \hline
  \end{tabular}
  \end{center}
  \item We then apply the external $CZ$ gates that connect qubits between different blocks. Using the approach of earlier sections, to maintain a separable decomposition, we allow each individual qubit radius to grow according to the number of external $CZ$ gates applied to that qubit. On a given block $b$ containing $n$ qubits, we tentatively define the block state space as the product $S'_b(r) := \bigotimes_{i=1,..,n} {\rm Cyl}(r_i)$, where for qubit $i$ in the block, $r_i=r \lambda^{e_i}$, where $e_i$ denotes the number of external $CZ$s the qubit undergoes. This means that the block state spaces are parameterised by the single parameter $r$. By construction, the resulting state (without yet having applied the internal $CZ$s) is separable with respect to these state spaces. Although there may be more complicated state spaces that give better eventual classical algorithms, by following the path we have taken we avoid the need for a potentially difficult separability analysis.

      So, in our $8 \times 8$ example, for instance, we will now have new cylindrical state spaces as follows
         \begin{center}
  \begin{tabular}{|cccc|cccc|}
    \hline
  $\cdot$ & $\cdot$ & $\cdot$ & $\circ$     & $\circ$ & $\cdot$ & $\cdot$ & $\cdot$  \\
  $\cdot$ & $\cdot$ & $\cdot$ & $\circ$     & $\circ$ & $\cdot$ & $\cdot$ & $\cdot$   \\
  $\cdot$ & $\cdot$ & $\cdot$ & $\circ$    & $\circ$ & $\cdot$ & $\cdot$ & $\cdot$  \\
  $\circ$ & $\circ$ & $\circ$ & $\bigcirc$     & $\bigcirc$ & $\circ$ & $\circ$ & $\circ$  \\
    \hline
   $\circ$ & $\circ$ & $\circ$ & $\bigcirc$     & $\bigcirc$ & $\circ$ & $\circ$ & $\circ$  \\
      $\cdot$ & $\cdot$ & $\cdot$ & $\circ$    & $\circ$ & $\cdot$ & $\cdot$ & $\cdot$   \\
      $\cdot$ & $\cdot$ & $\cdot$ & $\circ$    & $\circ$ & $\cdot$ & $\cdot$ & $\cdot$  \\
      $\cdot$ & $\cdot$ & $\cdot$ & $\circ$     & $\circ$ & $\cdot$ & $\cdot$ & $\cdot$   \\
    \hline
  \end{tabular}
  \end{center}
  In this diagram the dots represent ${\rm Cyl}(r)$, the smaller circles represent ${\rm Cyl}(\lambda r)$, and the biggest circles represents ${\rm Cyl}(\lambda^2 r)$.
  \item  We then obtain new state spaces $S''_b(r)$ by applying the internal $CZ$s to the $S'_b(r)$ state spaces we had in the previous step. This leads to a constraint on $r$, as for the classical simulation we require the state spaces $S''_b(r)$ to be contained within the dual of the permitted cluster circuit measurements. Let $r_{b,max}$ be the maximum value of $r$ such that $S''_b(r)$ is contained within the dual, and define our final state spaces as $S_b := S''_b(r_{b,max})$. For inputs with radius less than $r_{b,max}$, this furnishes a separable decomposition in terms of the block state spaces $S_b$, which can then be used to provide a classical efficient simulation.
\end{enumerate}
This coarse graining process can only increase the range of systems that we can efficiently simulate classically. To see why, let us compare the constraints we will have on $r$ from this coarse grained approach, to those obtained in the `fine grained' approach of earlier sections. In fact the only real difference between the approaches occurs in the last step. In the fine grained analysis, $r$ had to be picked so that when the internal $CZ$s are applied, not only was the block state space in the dual of the measurements, but the output state {\it also} had to be separable with respect to an internal partition into ${\rm Cyl}(1)$ spaces. This is a stronger requirement than simply requiring the internal $CZ$s to keep the block state space in the dual, and therefore leads to the possibility that coarse graining could allow us to simulate larger values of $r$. We will see that this is indeed the case, and in fact we will see that coarse graining into larger and larger blocks can only increase the values of $r$ that can be efficiently simulated classically.

\subsection{Blocks of 2 cylinders}

To illustrate the above approach, let us first consider a 2-dimensional square lattice, and suppose that the length on one side is even. Let us consider a partioning as follows into blocks of two particles, as illustrated here:
  \begin{center}
  \begin{tabular}{|c|c|c|c|c|c|c|c|}
    \hline
  $\cdot$ & $\cdot$ & $\cdot$ & $\cdot$ & $\cdot$ & $\cdot$ & $\cdot$ & $\cdot$  \\
 $\cdot$ & $\cdot$ & $\cdot$ & $\cdot$ & $\cdot$ & $\cdot$ & $\cdot$ & $\cdot$   \\
 \hline
$\cdot$ & $\cdot$ & $\cdot$ & $\cdot$ & $\cdot$ & $\cdot$ & $\cdot$ & $\cdot$  \\
 $\cdot$ & $\cdot$ & $\cdot$ & $\cdot$ & $\cdot$ & $\cdot$ & $\cdot$ & $\cdot$   \\
 \hline
   $\cdot$ & $\cdot$ & $\cdot$ & $\cdot$ & $\cdot$ & $\cdot$ & $\cdot$ & $\cdot$  \\
 $\cdot$ & $\cdot$ & $\cdot$ & $\cdot$ & $\cdot$ & $\cdot$ & $\cdot$ & $\cdot$   \\
 \hline
$\cdot$ & $\cdot$ & $\cdot$ & $\cdot$ & $\cdot$ & $\cdot$ & $\cdot$ & $\cdot$  \\
 $\cdot$ & $\cdot$ & $\cdot$ & $\cdot$ & $\cdot$ & $\cdot$ & $\cdot$ & $\cdot$   \\
    \hline
  \end{tabular}
  \end{center}
  We will call any such block of two qubits a `2block'.
Each qubit in a block away from the perimeter of the entire grid undergoes three external $CZ$s, and so the starting $r$ will be taken to
\begin{equation}
  r' := r \lambda^3
\end{equation}
for all particles not on the perimeter (the qubits on the perimeter would only grow by $\lambda$ or $\lambda^2$, but we won't consider them as
they will lead to a weaker constraint on $r$).
Previously, in the fine grained analysis, we would at this step apply the remaining $CZ$ (the internal one) between the qubits inside a given block, and that would lead to a constraint that
\begin{equation}
   r' \leq {1 \over \lambda} \Rightarrow r \leq {1 \over \lambda^4}.
\end{equation}
However, we now instead only need to make sure that we do not get taken out of the dual.
We will shortly show that $r_{2block,max}=1/2$, and so provided that
\begin{equation}
  r' \leq {1 \over 2} \Rightarrow   r \leq {1 \over 2\lambda^3}
\end{equation}
applying the internal $CZ$ does not take us out of the dual of the measurements we are permitted, and we have a useable separable decomposition
over blocks of 2 particles.
Hence by considering blocks of two qubits we obtain an increase, albeit slight, in the size of the region that can be efficiently simulated classically.

To see where $r_{2block,max} = 1/2$ comes from, we must consider all extremal inputs to the internal $CZ$ of the form:
\begin{multline}\label{input}
%\begin{equation} \label{input}
[1,r' \cos(\theta_A), r' \sin(\theta_A), \pm 1] \\
\otimes [1,r' \cos(\theta_B), r' \sin(\theta_B), \pm 1]
%\end{equation}
\end{multline}

and compute the maximum value of $r'$ such that for any inputs of this form and any allowed measurements, we obtain non-negative probabilities after the $CZ$ acts.

We first show that we can make a number of simplifications that reduce the inputs that we need to consider. These simplifications will be also needed in further coarse graining later:

\begin{enumerate}
\item {\bf W.l.o.g. we need only consider measurements in the $XY$ plane}. If either of the measurements on the two particles is a $Z$ measurement, no negativity can arise because if we (by cyclicity of the trace) act the $CZ$ on the measurement operators we simply
rotate the non-$Z$ measurement operator by $Z^0 = I$ or $Z^1=Z$. This means that the measurement could be replaced by a product measurement on products of cylinders, and no negativity could arise.
\item {\bf W.l.o.g. we need only consider inputs with $z=1$}. Suppose that we start with inputs that have $z=-1$ on a given cylinder. This can be described as a $z=+1$ input with an $X$ applied to it. However, pulling the $X$ through the $CZ$ (by a standard stabilizer calculation) to apply it to the measurement operators, gives an $X$ on the measurement for that cylinder, and $Z$s on measurements on the neighbours. These are just new $XY$ plane measurements, so a $z=-1$ can be transformed to a $z=+1$ simply by changing the $XY$ plane measurements being considered.
\item {\bf W.l.o.g. we may restrict to measurements projectors:}
\[{I - X \over 2}
\]
The reason for this is by the $Z$ symmetry of the cylinders and the $CZ$ gates, we may simply apply a $z$-axis rotation to any $XY$ plane projector to it into the projector $(I-X)/2$.
\end{enumerate}
Hence we now need to compute the maximal value of $r$ such that extremal cylinder inputs with that radius and $z=+1$ do not give negative overlap with projectors
\[{I - X \over 2} \otimes {I - X \over 2}
\]
The probability of getting this outcome on inputs satisfying the restrictions can easily be computed to be (up to an unimportant positive factor):
\begin{align}
%\begin{eqnarray}
1 - r' \cos(\theta_A) - r' \cos(\theta_B) + r'^2 \sin(\theta_A) \sin(\theta_B) \nonumber\\ 
= (1-r' \cos(\theta_A))(1-r' \cos(\theta_B)) - r'^2 \cos(\theta_A\theta_B) \label{2nd}
%\end{eqnarray}
\end{align}

The expression gives $1-2r'$ when $\theta_A=\theta_B=0$, so we need $r' \leq 1/2$. However, for any $0 \leq r' \leq 1/2$, this is the minimal possible value, as equation (\ref{2nd}) is no smaller than $(1-r')(1-r' ) - r'^2 = 1-2r'$. Hence the probability is positive for all measurements and all inputs iff $r' \leq 1/2 = r_{2block,max}$.

This example shows that coarse graining certainly helps to increase the size of the classical region, but as we now discuss one can do better by increasing the size of the blocks.

\subsection{Larger Blocks}

We will define two sequences of optimisation problems which bound each other, and are obtained by considering increasing block sizes.
The upper sequence is non-increasing and the lower sequence is non-decreasing. The limit
of the lower sequence gives the radius of inputs that can be efficiently simulated using the coarse graining approach described above. In the case of a 2D lattice
the two sequences converge to limits that are quite close, but we do not yet know whether the limits are the same for both sequences.
The basic principles can apply to other lattices that can be split into tiles in an appropriate way.

We begin by describing the two sequences. On any given rectangular block $B$ with $H \times W$ qubits (where $H,W \geq 2$) embedded in a larger lattice, we consider two ways of initialising the qubits:

\begin{itemize}
\item[i)] All cylinders are prepared in arbitrary extremal cylinder states with radius $r$ and $z=+1$, \,\,\, or

\item[ii)] All cylinders on interior qubits are prepared in arbitrary extremal cylinder states with radius $r$ and $z=+1$, but qubits in the boundary of the rectangle are prepared in extremal cylinder states with radius grown according to the number of external $CZ$s. So corner particles are prepared in extremal cylinder states with radius $\lambda^2 r$ and $z=+1$, and all other boundary qubits prepared in extremal cylinder states with radius $\lambda r$ and $z=+1$.
\end{itemize}

Consider applying internal $CZ$s to the block, and denote the resulting operators describing the whole $H \times W$ block by $\rho(B,r)$ and $\rho_{\lambda}(B,r)$ respectively. We will be interested in when these operators are in the dual of the permitted measurements. If an operator $\rho$ is in the dual of the set of permitted measurements $\mathcal{M}$ we will write $\rho \geq_{\mathcal{M}}0$ (reflecting the fact that the operator is `positive' with respect to the allowed measurements). We define the following quantities:
\begin{eqnarray} \label{optimum}
  s(B) &:=& \max \{ r | \rho(B,r) \geq_{\mathcal{M}} 0  \} \\
  s_{\lambda}(B) &:=& \max \{ r | \rho_{\lambda}(B,r) \geq_{\mathcal{M}} 0   \}
\end{eqnarray}
We are interested in the value of $s_{\lambda}(B)$ for increasing block sizes. For a given value of $r$, if there is a block $B$ such that $r \leq s_{\lambda}(B)$, then inputs from cylinders of radii $r$ can be efficiently simulated classically. On the other hand, if there is a block
size $B$ such that $r > s(B)$, then inputs from cylinders of radii $r$ lead to negative probabilities, and so there cannot be a separable decomposition
(which by construction leads to positive probabilities), and moreover the sampling problem would not be well defined in the first place for these values of $r$.

So, we are interested in finding blocks such that $s_{\lambda}(B)$ is large. The following lemma helps somewhat with this task.

{\bf Lemma 4:} Consider a region $KL$ of qubits embedded in a larger lattice. Consider cutting the region into two disjoint subregions $K$ and $L$ (i.e. we remove the $CZ$ gates joining these two regions). Then we have the following relationships. For any region $F$ whatsoever:
\begin{equation} \label{upperseq}
 s(F) \geq s_{\lambda}(F)
\end{equation}
For the region $KL$ and its subregions $K$ and $L$:
\begin{eqnarray}
  s(KL) &\leq& \min \{s(K),s(L) \} \label{decrease} \\
  s_{\lambda}(KL) &\geq& \min\{ s_{\lambda}(K),s_{\lambda}(L)\} \label{increase}
\end{eqnarray}
In particular suppose that the two subregions $K$ and $L$ are identical (i.e. correspond to isomorphic graphs), then in that case we would have (writing $K=L$ to denote that the subregions are isomorphic):
\begin{eqnarray}
  s(LL) &\leq& s(L) \\
  s_{\lambda}(LL) &\geq& s_{\lambda}(L)
\end{eqnarray}
Informally this tells us that as we increase the size of a region by repeatedly joining subregions together, then $s$ can only decrease, whereas $s_{\lambda}$ can only increase, in spite of the fact that they are are defined via similar looking optimisation problems.

\medskip

\noindent {\bf Proof:} To see equation (\ref{upperseq}) note that $\rho(F,r)$ can be obtained from $\rho_{\lambda}(F,r)$ by dephasing the externally connected qubits of $F$. As dephasing maintains $\geq_{\mathcal{M}} 0$ positivity, this means that $ s(F)\geq s_{\lambda}(F)$.

To see equation (\ref{decrease}) we observe that cutting the $CZ$ gates joining regions $K,L$ does not change the marginal operators on $K$ or $L$, i.e. $\tr_K\{ \rho(KL,r)\} = \rho(L,r) $ (and as the labelling of the region is not important, this is true if we interchange $K$ and $L$ too). This may be seen by a simple computation: a cylinder extremum with $z=+1$ can be written in the form $\ket{0}\bra{0}+ a\ket{0}\bra{1} + b\ket{1}\bra{0}$ for
some $a,b \in \mathbb{C}$. Consider for instance interacting this with two other particles in an arbitrary state $T$, using two $CZ$ gates. They become:
\begin{equation*}
\ket{0}\bra{0} \otimes T + a\ket{0}\bra{1} \otimes (T Z^{\otimes 2}) + b\ket{1}\bra{0} \otimes (Z^{\otimes 2} T)
\end{equation*}
Tracing out the first particle leaves the remaining particles in their original marginal state $T$. One can see that this would be the case irrespective of the number of $CZ$ gates applied. Hence the marginal state of a given region does not change when external $CZ$s are applied. Now, if $\rho(KL,r)\geq_{\mathcal{M}}0$ then we also have $\tr_K \{\rho(KL,r)\}\geq_{\mathcal{M}}0$, but as $\tr_K\{\rho(KL,r)\}=\rho(L,r)$ this means that $\rho(KL,r)\geq_{\mathcal{M}}0$ implies both $\rho(K,r)\geq_{\mathcal{M}} 0$ and $\rho(L,r)\geq_{\mathcal{M}}0$, hence we have equation (\ref{decrease}).

To see equation (\ref{increase}), we note that $\rho_{\lambda}(KL,r)$ is in the convex hull of products $\rho_{\lambda}(K,r) \otimes \rho_{\lambda}(L,r)$ by using the stochastic representation of the $CZ$s that join $K$ and $L$. Hence $\rho_{\lambda}(KL,r)$ on the block $KL$ must be $\geq_{\mathcal{M}} 0$ if $\rho_{\lambda}(K,r)\geq_{\mathcal{M}}0$ and $\rho_{\lambda}(L,r)\geq_{\mathcal{M}}0$, and so equation (\ref{increase}) must hold. $\blacksquare$

We remark that Lemma 4 also applies to privileged basis architectures, the only argument that needs to be adjusted is the marginal argument, see footnote \footnote{All the arguments as presented for the qubit case go through unchanged for privileged basis architectures, except for the argument about marginals, which although very similar may require slight clarification. In privileged basis architectures the cylinder on a single particle will have extremal operators of the form $\sigma = |a\rangle \langle a|+ \Delta$ where $\Delta$ represents off-diagonal terms. Consider a controlled-diagonal unitary of the form $U = \sum |i\rangle \langle i| \otimes Z_i$, where $Z_i$ represents a diagonal single particle gate contingent on the the value of $i$. We note that $U \sigma \otimes \sigma' U^{\dag} = \sum_{a,b}  \bra{a} \sigma \ket{b}  \ket{a} \bra{b} \otimes Z_a \sigma' Z^{\dag}_b $. Tracing out the first particle leaves the second particle in $Z_a \sigma' Z^{\dag}_a$ for one specific value of $a$ when $\sigma$ is an extremal point, hence the marginal state on the second particle is at most transformed by a diagonal single particle unitary. So we see that when extremal points are considered, $U$ interactions and then tracing out at most rotates the other qubits by diagonal unitaries. As a diagonal unitary on a given particle does not change its positivity with respect to the allowed measurements, we have that the positivity of a state implies the positivity of any marginal, so the analogue of $u_n$ for these systems will also be non-increasing.} for an explanation.

Lemma 4 allows us to define sequences that help to capture when $r$ is classically simulatable efficiently through the coarse graining approach.
Consider for instance constructing a sequence of blocks by starting with a single $2 \times 2$ block $B_1$ and then recursively constructing larger blocks by joining two copies of $B_{n-1}$ to make $B_n$.
Define sequences
\begin{eqnarray}
  u_n &:=& s(B_n) \\
  l_n &:=& s_{\lambda}(B_n)
\end{eqnarray}
From Lemma 4 we have that $u_n \geq l_n$, $l_n$ is non-decreasing, and $u_n$ is non-increasing, and hence both sequences converge. Let us denote the limits as:
\begin{eqnarray}
  u &:=&   \lim u_n \\
  l &:=&  \lim l_n
\end{eqnarray}
A radius $r$ is classically simulatable efficiently if $r < l$ but if $r > u$ then it leads to negative probabilities, and so in the latter situation the problem of classically sampling from the output of the `cylindrical computer' is not well defined.
It is natural to speculate that $u=l$. This would have an interesting foundational interpretation: if we had a hypothetical cylindrical computer made from cylindrical bits placed on the vertices of the lattice and undergoing $CZ$ interactions with their neighbours, then for $r < l$ the system would be efficiently simulatable classically, and for $r>l$ the system would not give valid probabilities. A similar interpretation would hold for any other privileged basis architecture and lattice if they have $u=l$.

\begin{figure}[ht!]
\centering
%\hspace{-1cm}
\includegraphics[width=80mm]{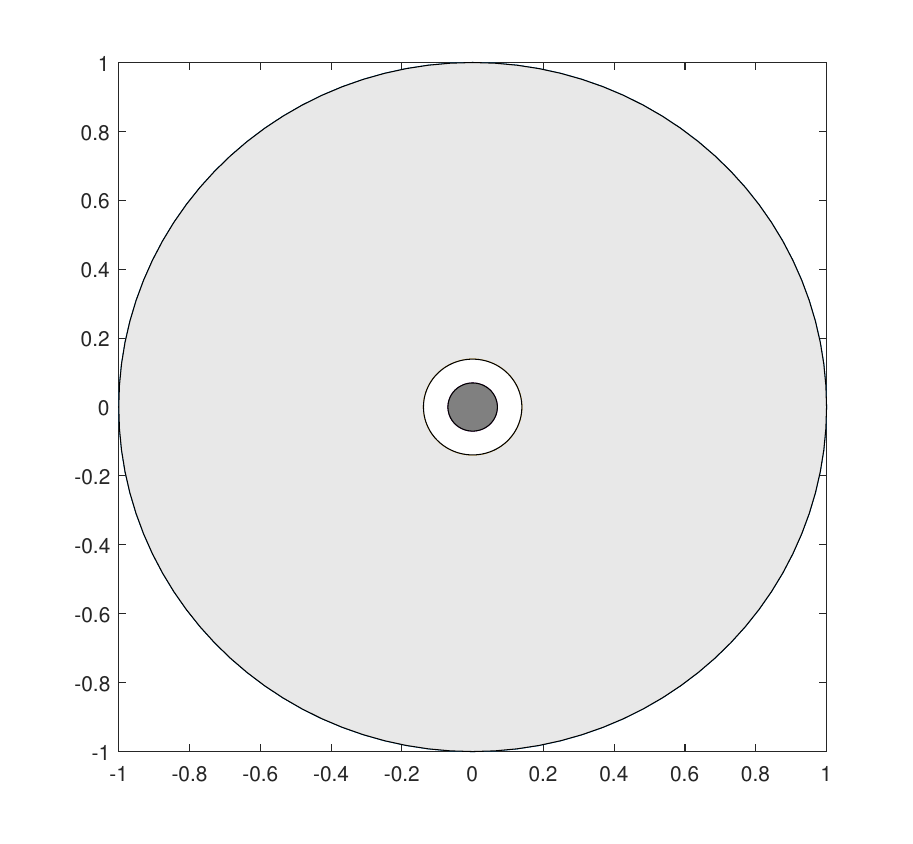}
\caption{We define `cylindrical computers' to be hypothetical devices made by placing `cylindrical bit' operators $(I+r\cos{\theta}X+r\sin(\theta)Y \pm Z)/2$, taken from the top or bottom of a cylinder
of radius $r$, at the vertices of a lattice, interacting neighbours with $CZ$ gates, and then measuring in $Z$ or $XY$ plane measurements. The diagram represents the input operators (of either $z$ value) as points $(r,\theta)$ in polar coordinates. In the case of a square 2D lattice, we know that inputs from the dark grey central region $r < 0.0698$ can
be efficiently simulated classically by the coarse graining approach, whereas inputs from the outer lighter grey region $r > 0.139$ lead to negative probabilities (given a large
enough lattice). We are certain that these bounds are not tight. The narrow white band represents the currently uncertain region - computing more terms in the sequences $u_n$ and $l_n$ will make this region narrower. If the limits $u$ and $l$ meet, then the white band would shrink to a circle at radius
$l$. Diagrams of a similar nature can in principle (although the computations may be difficult) be produced for any privileged basis architecture with inputs from its cylinders.}
 \label{bullseye}
\end{figure}

While we have not been able to establish whether or not $u=l$ for any system, for a square 2D lattice with $CZ$ interactions we have numerically computed lower bounds on $l_1$ (using a polyhedral outer approximation of the input cylinders and doing a brute force search), and upper bounds to $u$ using  trial measurements and inputs on rectangles of size $6 \times 7$. These numbers indicate that for a 2D square lattice $0.0698 \leq l \leq u \leq 0.139$ (but tentative further investigations in fact suggest that $0.0913 \leq l \leq u \leq 0.128$). Hence even if $u \neq l$, they are not far apart (see figure \ref{bullseye}).

These numerical investigations can certainly be taken further. We leave this for another occasion. However, we will report one initial finding: for the square 2D lattice, numerical experiments seem to suggest the following conjecture: both the upper sequence $u_n$ and the lower sequence $l_n$ are determined by considering measurement projectors on each particle of the form $(I-X)/2$, and input extrema of the form $(I+ \alpha X +Z)/2$ (i.e. with no $Y$ component) where $\alpha$ includes the contribution from $r$ and any growth factors $\lambda$ applied in the coarse graining process. The maximum $r$ for which these inputs and measurements give positive probability appears to be the maximum in equations (\ref{optimum}).

%%%%%%%%%%%%%%%%%%%%%%%%%%%%%%%%%%%%%%%%%%%%%%%%%%%%%%%%%%%%%%%%%%%%%%%%%%%%%%%

\section{Obstacles to classical simulation}  \label{SectionObstacles}

In the foregoing discussion we discussed our classical simulation methods in the context of input state spaces that are cylinders or coarse grained versions of them, and hence contain many non-physical operators.
However, as we are ultimately only interested in simulating systems with {\it quantum} inputs, we may wonder whether it is possible to change our state spaces (e.g. perhaps shaving off
some extremal points) to obtain a greater range of quantum inputs that can be efficiently simulated classically. Or more generally, one might wonder whether there is some other classical simulation algorithm that
can simulate more quantum inputs (e.g. with a higher radius, but contained within the Bloch sphere). In this section we discuss obstacles that such endeavours may face. 

The main ideas are already present in other works  \cite{Terry,mora_universal_2010}, we merely modify them slightly to fit our setting in which we have stronger restrictions on the available measurements, as we are not permitted to remeasure qubits. The key observation is that if we start with input qubits with high enough radius, then we can use the cluster circuits to steer unmeasured qubits to a $\ket{+}$ state conditioned on the measurement outcomes of permitted measurements \footnote{We thank Miguel Navascues and Richard Jozsa for suggesting this line of investigation to us.}. This means that one could, given qubits of a high enough radius on an appropriate lattice, probabilistically prepare in ideal cluster state on the unmeasured qubits. This could in turn give obstacles to classical simulation algorithms. We consider two ways:
\begin{enumerate}
\item Following arguments similar to those utilised in \cite{Terry,mora_universal_2010}, if the probability of success for creating a $\ket{+}$ on unmeasured qubits exceeds a threshold determined by lattice percolation thresholds on the unmeasured lattice, one can implement cluster state quantum computation. This means that for some graphs of finite degree, there is a radius beyond which BQP can be supported and classically efficient simulation is unlikely.
\item One could rule out the existence of a separable decomposition under any coarse graining scheme using the fact that one can violate a Bell inequality using the permitted measurements. As separable decompositions automatically furnish a local hidden variable model \cite{Werner}, non-locality would rule out a separable decomposition based classical algorithm, even if not ruling out other classical algorithms.
\end{enumerate}
In appendix \ref{appendix} we present an example of the first of these approaches: on a lattice of degree $5$ (see figure \ref{fig:lattice}) with input {\it quantum} pure states drawn from within ${\rm Cyl}(r_{max})$ with $r_{max}=0.84$ one can create a perfect 2D cluster state efficiently on one subset of the qubits by measuring the other qubits. Hence it should not be possible to classically efficiently simulate quantum pure states with $r \geq 0.84$ on such a lattice. 

The nonlocality obstructions are essentially questions of localisable `non-locality', in a similar sense to the definition of localisable entanglement \cite{localisable}. Imagine that we are attempting to find a separable decomposition, with any state space or coarse graining method, such that two adjacent regions of qubit are in different blocks, across which we would like a separable decomposition. As the regions are adjacent there will be one qubit in one block connected to another qubit in another block by a $CZ$ gate. Pick two such qubits and call them $A$ and $B$. Mark out two chains of qubits, one from the first block terminating at $A$, and the other from the second block terminating at $B$. If we measure out all qubits in the $Z$ basis, except for ones on the two chains, assuming that the initial $r$ was high enough, one can use a similar protocol to appendix \ref{appendix} to create $\ket{+}$ states in $A$ and $B$. This will result in an EPR pair that can then violate a Bell inequality with our permitted measurements, and so no separable decomposition can be used as soon as the input radii are high enough for this purification to be possible. 
This means that when the initial radius $r$ is too high, there can be no suitable generalised separable decomposition, even with a different choice of state space. For small values of $r$ a similar process would localise {\it quantum} entanglement between the two qubits, but not non-locality for our restricted measurements.

Further obstacles to increasing the set of inputs for which we can classically simulate might be obtainable from conjectures about the polynomial hierarchy. In these arguments \cite{harrow_quantum_2017}, one proves that if widely-believed complexity theoretic conjectures hold (i.e. the non-collapse of the polynomial hierarchy) then there cannot be any efficient classical simulation algorithm.

For example, one can entertain the possibility of a multiplicative error simulation. To prove that our restricted model on a 2D lattice cannot be simulated, we could consider attaching linear chains of ancilla qubits to the qubits on the lattice. We would then want to show that by measuring the ancillas, and allowing for post-selection, we can prepare a state on the lattice that is universal resource state for post-selected MBQC. Then by following the arguments used in \cite{bremner_classical_2010}, this suffices to show that the restricted 2D cluster state cannot be simulated up to multiplicative error. However, this notion of simulation is physically unrealistic, and we would like to rule out a classical simulation up to additive error. Stemming from work by \cite{bremner_average-case_2016,aaronson_complexity-theoretic_2016}, there has been further progress in ruling out additive error simulations for various restricted models of quantum computing (\cite{bremner_achieving_2017,miller_quantum_2017,gao_quantum_2017,yoganathan_quantum_2019,haferkamp_closing_2020}). For now however, we leave this for future work.

Nevertheless, we note here that when post-selection is permitted, allowing all measurements (not just $Z$ or $XY$ plane measurements) brings additional power. Consider two qubits with a low radius $r$ undergoing a $CZ$ gate. The existence of a cylinder separable decomposition for the output shows that if we then post-select on on the outcomes of $Z$ or $XY$ plane measurements on one qubit, we cannot steer the unmeasured qubit to a perfect $\ket{+}$ (as the other qubit must be taken to a state from inside a cylinder of radius $\lambda r$). However, if we are permitted to measure the first qubit arbitrarily, then  (by standard considerations \cite{HJW}) with postselection one can obtain a perfect $\ket{+}$ on the second qubit.

\section{Discussion} \label{SectionDiscussion}

We have shown that computations made from cluster state circuits acting upon inputs close enough to computational basis states can be efficiently simulated classically. We obtain explicit bounds in the case of the qubit systems, but the framework applies to qudits and other types of (diagonal) interaction as well. Our classical simulations also lead to two types of local hidden variable model, the second of which is non-standard, as the hidden variable model can communicate within blocks. The inital classical simulations furnish examples of highly entangled quantum systems that have a local hidden variable model through the cylinder separable decomposition. The second coarse graining simulations lead to a kind of local hidden variable model in which the locality constraint is relaxed for particles within the same block.

Let us offer some remarks on how the approach given in this work could probably be applied in other situations. Key to our construction has been the idea that to maintain a separable decomposition one can grow the local state space to include non-physical operators. However, this is completely general: given any gate (diagonal or not) one can maintain a separable decomposition by `growing' the state space. That this is true is actually just a different take on standard ideas in entanglement theory. For instance consider the fact an entangled state can be turned into a quantum-separable one by local noise acting upon each particle. This means that acting upon the quantum-separable state with the inverse of the noise will give us a separable decomposition for the original entangled state, albeit one involving state spaces that have `grown' larger than physical quantum ones. The problem with non-physical separable decompositions like this is that they cannot be sampled from as they lead to negative probabilities. However, if we are only interested in certain measurements, we could hope that any negativities that arise will be controlled well enough to not be `seen' by our permitted measurements. This is exactly what we have done in this work through the use of cylindrical state spaces. The fact that this can lead to classically efficient simulations of non-trivial pure quantum-entangled systems is perhaps surprising, and it suggests that in other situations, perhaps other low-degree circuits with more general (non-diagonal) gates, the approach could be more effective than might be anticipated.

An important consideration in any such investigations would be how quickly the state spaces must grow. In the case of two particle states, the construction of `small' state spaces that provide a separable decomposition with minimal `negativity' has been considered in \cite{AJRV2_1,AJRV2_2}, and connections to cross norm entanglement measures \cite{Oliver} provide a useful technical tool for attempting to minimise state space growth. The approach is closely related to the general theme of using quasi-probability distributions for classical simulation methods and local hidden variable models. It is hence reasonable to expect that recent works that have explored simulations involving small amounts of `magic' or `negativity' (see e.g. \cite{seddon2021quantifying,Pashayan2015WB}) could also be combined with the notion of generalised-separability to simulate systems with small amounts of generalised entanglement.

In any given system it is possible that a different choice of state spaces could lead to classical simulation algorithms for other quantum inputs. For example, in the context of the cluster state variants that we have considered in this work, we do not directly care about simulating systems with non-physical cylindrical inputs, we only care about quantum inputs. So we might consider other state spaces of different shapes, with the aim of finding ones that grow slowest when undergoing interactions, but contain as many quantum input states as possible. To what extent might this be possible? A convex hull version of twirling \cite{Werner} can be used to argue that an optimal state space must respect the symmetry group, and as pointed out earlier there are strong connections to certain entanglement measures \cite{AJRV2_1,AJRV2_2}. However in future work it may be useful to develop better systematic methods of constructing good state spaces. 

Another possible question is whether coarse graining can increase the class of permitted measurements that can be efficiently classically simulated. In this work we only used coarse graining to increase the set of initial inputs that could be simulated. However, it is possible that coarse graining could instead increase the set of simulatable measurements. For example, if it were possible to write down a non-trivial family of entangled states that are separable with respect to block state spaces consisting of entanglement witnesses \cite{witness}, then these states would be good candidates for entangled systems that can be efficiently simulated classically for {\it any} single particle measurements. This is because the set of entanglement witnesses can be considered to be non-quantum states spaces that are (by definition) in the dual of local measurements. We do not know if such examples exist.

Another viewpoint of the work is through its connections to the foundations of quantum theory. We can view our investigations as an exploration of the complexity of a kind of toy non-physical theory in which extremal cylinder operators (which are not quantum states) are placed at the nodes of a lattice, interacted with diagonal gates, and measured in a computational basis or in bases unbiased to it. In its present form this `theory' leads to negative probabilities for lattices of high enough degree, and so does not immediately  make `operational sense' as a physical theory. However, models of computation incorporating quasi-probability considerations have been considered in the work of \cite{Lee}, and we currently believe that the cylindrical computers that we consider here could be considered operational `non-free' theories according to the definition of those authors. While our focus has been on using cylindrical computers with $r < l$ for constructing classical simulation algorithms for quantum systems, the framework of \cite{Lee} could shed light on the case when $r > u$. We also note that the systems we consider could lead to an interesting dynamical theory from a field theory perspective, where for a regular square lattice one direction could represent time (as is a standard interpretation in cluster state computation). 

Another open question is whether the limits $u$ and $l$ are identical for some lattices that are amenable to coarse graining. If it turns out to be the case that $u=l$ for such a system, then that would mean that apart from $r=l$ we would know the computational power almost entirely - for $r>l$ the system gives negative probabilities (and so the sampling problem is not well defined), but for $r<l$ the system is classically tractable. In the case of the 2D square lattice of qubits with $CZ$ interactions, we found that $u$ and $l$ are certainly close, and this difference can be made smaller still by computing more terms in the sequences.

Although the focus of our work has been classical simulation, any attempts to optimise state spaces in generalised separable decompositions, or even classically simulate in any way, will eventually face obstacles coming from computational complexity conjectures. We have seen that following the approach of \cite{Terry,Dan}, if we take quantum mechanical inputs to cluster circuits, then the state of one particle may increase in radius conditioned on the outcome of measurements elsewhere, and indeed for some lattices we can steer qubits into $\ket{+}$ states. If our starting qubits have high enough radius, this can happen with sufficiently high probability that $BQP$ can be recovered using the percolation style arguments of \cite{Terry,Dan}. This means that there are explicit connections between our results and percolation thresholds assuming that $BQP \neq BPP$, as input values of $r$ that are suffficiently high for quantum computation via percolation methods must be higher than the values of $r$ than we can efficiently simulate classically. The ability to simulate $BQP$ through postselection could also in principle be the starting point of polynomial hierarchy obstacles \cite{TerhalD,harrow_quantum_2017,bremner_classical_2010,bremner_average-case_2016,aaronson_complexity-theoretic_2016}  to classical simulation which could potentially apply for lower values of $r$ and lower degree. \newline

\section{Acknowledgements}

We thank David Gross, Richard Jozsa, Akimasa Miyake, Ashley Montanaro, and Miguel Navascues for helpful conversations at various stages of the work.
At various stages this work was supported a University of Strathclyde starter grant (Virmani), an Imperial College Junior Research fellowship (Jevtic), EPSRC grant EP/K022512/1 (Jevtic, Virmani), and
through an EPSRC DTP award (Garn). Michael Garn gratefully acknowledges the support of a Prachi Dwivedi award.

\appendix

\section{Stochastic representation} \label{AppA}

To construct the stochastic representation let us initially assume we are only interested in input states with $z=1$, we will consider other values of $z$ shortly. Given an input radius $r$, define a standard `fiducial' extremal input with $z=1$, say $\rho_0(r) := (I + rX +Z)/2$
(it doesn't really matter which we pick).
As any cylinder extremum with $z=1$ can be reached from a fiducial state by the action of a local $Z$ rotation, we can write the output separable decomposition in terms of fiducial states:
\begin{equation} \label{fiducial}
  CZ(\rho_0(r) \otimes \rho_0(r)) = \sum_{i=1}^{K} p_i U_i \rho_0(\lambda r) U^{\dag}_i \otimes V_i \rho_0(\lambda r) V^{\dag}_i
\end{equation}
where $U_i$ and $V_i$ are local $Z$ rotations, and $p_i$ is a probability distribution. Now if we want to construct the action of the $CZ$ on two other input states $\sigma \otimes \omega$ (with $z=1$) we may simply express these new inputs
in terms of the fiducial states, $\sigma \otimes \omega = S\rho_0 S^{\dag} \otimes W \rho_0 W^{\dag}$ for two $Z$ diagonal unitaries $S,W$, and apply the same separable decomposition because everything commutes with local $Z$ rotations:
\begin{equation}
  CZ( \sigma \otimes \omega ) = \sum_{i=1}^{K} p_i U_i S \rho_0(\lambda r) S^{\dag} U^{\dag}_i \otimes V_i W \rho_0(\lambda r) W^{\dag} V^{\dag}_i
\end{equation}
This means that the action of the $CZ$, for inputs with $z=1$, can be represented by a radius growth by $\lambda$ and the ensemble of unitaries $\{p_i,U_i \otimes V_i\}$. We must now consider what happens if any of the inputs has $z=-1$. This can be accounted for in the separable decomposition (\ref{fiducial}) by changing all $z$ values to $-1$ for the inputs with $z=-1$, and applying a $Z$ rotation to the other particle. Hence for any inputs we may represent
the $CZ$ by radius growth, the action of $\{p_i,U_i \otimes V_i\}$, and possibly extra $Z$ rotations where the input states have $z=-1$ (note that such additional $z$ dependence is unavoidable, because the $CZ$ gate can communicate information from one particle to the other, and an operation $\{p_i,U_i \otimes V_i\}$ with growth of $r$ by itself does not communicate from one particle to the other).

\section{Purifying to $\ket{+}$ states within lattices} 

\label{appendix}

In this appendix we show that a lattice of degree $5$ with input {\it quantum} pure states drawn from within ${\rm Cyl}(r_{max})$ with $r_{max}=0.84$ can be converted to a perfect 2D cluster state. Our approach uses similar arguments as used in \cite{Terry}, in which a state on a lattice is prepared by applying CZ gates, on edges of the lattice, to a product state where each qubit is close to $\ket{+}$. Subsequently, a local 2-outcome measurement is applied to each qubit on the lattice, which either disentangles the qubit from the lattice or projects it into a $\ket{+}$ state. It is then known from \cite{Dan}, that if the site occupation probability on the lattice is above a threshold $p_c$, then the resulting cluster state with holes is a universal resource for MBQC. Note that similar ideas were also used in \cite{mora_universal_2010}. In our model however, we do not permit remeasuring of qubits and we have further restrictions on the permitted measurements (i.e. we only use $Z$ basis measurements and $XY$ plane measurements), so we have to use a minor modification of previous arguments.

\begin{figure}
\begin{tikzpicture}[scale=.85]
\draw[help lines, color=gray!30, dashed] (-1,-1) grid (9,6);

 \foreach \i in {0,2,4}
{
\fill (\i,0.5) circle (0.15cm);
\fill (\i,1) circle (0.15cm);
\fill (\i,1.5) circle (0.15cm);
\draw[black,thick,dashed] (\i,0) -- (\i,1.5); 
}
 
\foreach \i in {2,4,6}
{
\fill (\i,2.5) circle (0.15cm);
\fill (\i,3) circle (0.15cm);
\fill (\i,3.5) circle (0.15cm);
\draw[black,thick,dashed] (\i,2) -- (\i,3.5); 
}

\foreach \i in {4,6,8}
{
\fill (\i,4.5) circle (0.15cm);
\fill (\i,5) circle (0.15cm);
\fill (\i,5.5) circle (0.15cm);
\draw[black,thick,dashed] (\i,4) -- (\i,5.5); 

%draw lattice
\foreach \i in {0,2,4}
{
  
    %xshifts the horizontal line  1 place every loop 
    \draw[darkgray,xshift = \i cm] (-0.5,\i) -- (4.5,\i);
     %slants the vertical line 1 place every loop
     \draw[darkgray,xslant=tan 45] (\i,-0.5) -- (\i,4.5);
}

\foreach \i in {0,2,4}
{

     %slants the vertical line 1 place every loop
     \draw[darkgray,xslant=tan 45] (\i,-0.5) -- (\i,4.5);
}

\foreach \i in {0,2,4}
{
  \foreach \j in {0,2,4}
     {
    \fill[lightgray,xshift = \i cm] (\j,\i) circle (0.15cm);
    }

 }
 
}
\end{tikzpicture}
   \caption{This diagram illustrates how to prepare a single $\ket{+}$ state on the lattice via a linear chain. The linear chain, attached vertically to the 2D lattice, is built from ancilla qubits that are initialised with certain specified angles, which are then measured in the $X$-basis. In the method described in the text and an appendix, one linear chain is attached to each qubit on the $n\times m$ lattice.
          }
    \label{fig:lattice}
\end{figure}
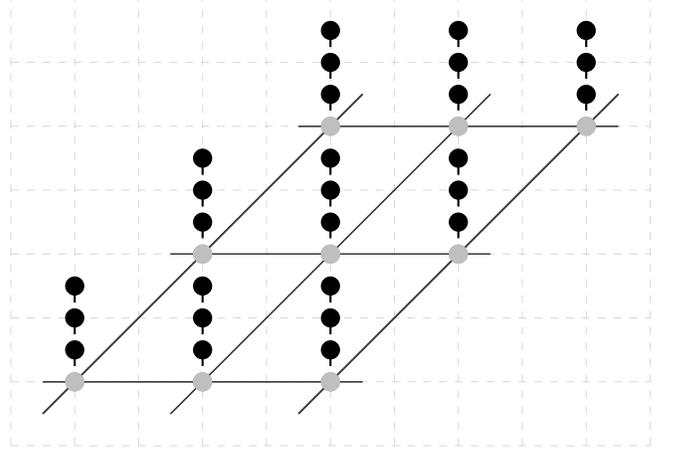

Instead of describing the inputs in terms of $r$ will now use a quantum pure state description, as the inputs are taken from the surface of the Bloch sphere.
The initial product state on the $n\times m$ lattice is
\begin{equation}
    \ket{\psi_{n\times m}} = \bigotimes_{i=1}^{N} \left(\cos(\phi_i/2)\ket{0} + \sin(\phi_i/2)\ket{1}\right),
\end{equation}
where the index $i$ denotes the qubit site, $N=n m $ is the total number of qubits on the lattice and $0 \leq \phi_i \leq \phi_{max}$. A CZ gate is then applied to each edge on the lattice. Note the correspondence between the radius and angle is given by $r=\left|\sin{\phi}\right|$. The fidelity with the usual perfect cluster state is then $\prod_{i}^{N} (\frac{1+\sin{\phi_i}}{2})$, and the perfect cluster state is recovered when $\phi_i = \pi/2$. 
If the input qubits were not $\ket{+}$ states, this is not an ideal cluster state. However, one can show that by attaching and measuring at most three ancilla qubits to each qubit (see figure \ref{fig:lattice}), we can probabilistically prepare $\ket{+}$ 2D cluster states suitable for quantum computation. The starting lattice (including the ancilla qubits) is hence the degree $5$ graph illustrated in figure \ref{fig:lattice}. To see how to to perform universal quantum computation, consider the following sequence of operations. 

\begin{enumerate}
    \item Prepare two ancilla qubits $\ket{\phi_1} $ and $ \ket{\phi_2}$, where $\ket{\phi_j} = \cos(\phi_j/2)\ket{0} + \sin(\phi_j/2)\ket{1}$, $\phi_j \in (0,2\pi)$ and the index $j$ denotes the qubit. Additionally, for technical reasons we impose the condition that $\phi_1 + \phi_2 = \frac{\pi}{2}$. 
    \item Apply a CZ gate between ancilla qubits 1 and 2 and measure the 1st qubit in the X-basis. If the outcome $x_1 = 1$ is obtained, which occurs with probability  $p_{x_1}(1) = \frac{1}{2}(1 - \sin{\phi_1}\cos{\phi_2})$, the post-measurement state of qubit 2 is 
        $$
\begin{aligned}
    \ket{\phi_2^\prime} =
    &\frac{\left[ \cos(\phi_1/2) - \sin(\phi_1/2)\right]\cos(\phi_2/2)}{\sqrt{1-\sin{\phi_1}\cos{\phi_2}}}\ket{0}
    \\
    + &\frac{\left[ \cos(\phi_1/2) + \sin(\phi_1/2)\right]\sin(\phi_2/2)}{\sqrt{1-\sin{\phi_1}\cos{\phi_2}}}\ket{1}.
\end{aligned}
$$
Therefore, if we pick $\phi_1 + \phi_2 = \frac{\pi}{2}$, the post-measurement state $\ket{\phi_2^\prime}$ becomes a $\ket{+}$ state.

\item If the outcome $x_1 = 0$ is obtained, which occurs with probability  $p_{x_1}(0) = \frac{1}{2}(1 + \sin{\phi_1}\cos{\phi_2})$, the post-measurement state of qubit 2 is

$$
\begin{aligned}
    \ket{\phi_2^\prime} =
    &\frac{\left[ \cos(\phi_1/2) + \sin(\phi_1/2)\right]\cos(\phi_2/2)}{\sqrt{1+\sin{\phi_1}\cos{\phi_2}}}\ket{0}
    \\
    + &\frac{\left[ \cos(\phi_1/2) - \sin(\phi_1/2)\right]\sin(\phi_2/2)}{\sqrt{1+\sin{\phi_1}\cos{\phi_2}}}\ket{1}.
\end{aligned}
$$

That is, $\ket{\phi_2}$ has undergone a rotation about the $Y$ axis toward the $\ket{0}$ state.
\end{enumerate}

If the outcome $x_1 = 1$ is obtained, then we have successfully produced a $\ket{+}$ state which is placed on the lattice. If the wrong outcome $x_1 = 0$ is obtained, then we initialise another ancilla qubit $\ket{\phi_3}$, such that $\phi_{2}^\prime + \phi_3 = \frac{\pi}{2}$, where $\phi_{2}^\prime$ is the angle of the post-measurement state of qubit 2. We then proceed to repeat the above procedure. That is, we apply a CZ gate between qubits 2 and 3, and measure qubit 2 in the $X$ basis. Similarly, if the outcome $x_2 = 1$ is obtained, with probability  $p_{x_2}(1) = \frac{1}{2}(1 - \sin{\phi^{\prime}_2}\cos{\phi_3})$, then the post-measurement state of qubit 3 is $\ket{+}$. If outcome $x_2 = 0$ is obtained, which occurs with probability  $p_{x_2}(0) = \frac{1}{2}(1 + \sin{\phi^{\prime}_2}\cos{\phi_3})$, then the post-measurement state of qubit 3 is $\ket{\phi^{\prime}_3}.$
By repeating this method, we can calculate the probability that the lattice site will be occupied by a $\ket{+}$ state. For example, repeating the method for three ancilla qubits, the probability is

\begin{equation}
     p_{site} = p_{x_1}( 1 ) +  p_{x_1}(0)\left[p_{x_2}( 1 ) + p_{x_2}( 0 ) p_{x_3}( 1 )\right] .
\end{equation}
In the case that a $\ket{+}$ has not been successfully prepared on the lattice by the ancilla chain, we measure the final qubit in the Z basis. This projects the qubit into the $\ket{0}$ (or $\ket{1}$) state which corresponds to creating a hole on the lattice, i.e. we have removed a vertex and edges from the cluster state. According to the percolation threshold $p_c = 0.5927\ldots$, if  $p_{site} > p_c$ then by \cite{Dan} we can construct an efficient LOCC algorithm that creates a perfect cluster state from a 2D cluster state with holes. We find that by attaching and measuring three ancilla qubits, with angles  $\phi_1 = 0.18\pi$, $\phi_2 = 0.32\pi$ and  $\phi_3 = 0.31\pi$,  we can prepare a $\ket{+}$ state on the lattice with probability $0.73$ which is above the percolation threshold $p_c$. The maximum angle required $\phi_2= 0.32\pi$ corresponds to $r_{max} = 0.84$. Therefore, we can prepare a $\ket{+}$ state with probability above the percolation threshold $p_c$, with three ancilla that are drawn from within ${\rm Cyl}(r_{max})$ with $r_{max} = 0.84$.

%%%%%%%%%%%%%%%%%%%%%%%%%%%%%%%%%%%%%%%%%%%%%%%%%%%%%%%%%%%%%%%%%%%%%%%%%%%%%%%
\bibliographystyle{plainnat}

%%%%%%%%%%%%%%%%%%%%%%%%%%%%%%%%%%%%%%%%%%%%%%%%%%%%%%%%%%%%%%%%%%%%%%%%%%%%%%%
\end{document}